\documentclass[twocolumn]{aastex701}

\usepackage{soul}
\usepackage[T1]{fontenc}
\usepackage{bm}

\shorttitle{Molecular Similarity and Water Diversity in Coeval Binary Disks}
\shortauthors{Yao et al.}
\journalinfo{The Astrophysical Journal, 1005:70 (19pp), 2026 July 1}
\received{2026 February 20}
\revised{2026 May 22}
\accepted{2026 May 24}
\published{2026 June 25}

\begin{document}

\title{Molecular Similarity and Water Diversity in Coeval Binary Disks: \\JWST/MIRI Observations of Sz 65 and Sz 66}

\correspondingauthor{Jinghuai Yao}
\email{jyao224@wisc.edu}

\author[0009-0002-0525-8222]{Jinghuai Yao}
\affiliation{Department of Astronomy, University of Wisconsin-Madison, Madison, WI 53706, USA}
\email{jyao224@wisc.edu}

\author[0000-0002-0661-7517]{Ke Zhang}
\affiliation{Department of Astronomy, University of Wisconsin-Madison, Madison, WI 53706, USA}
\email{ke.zhang@wisc.edu}

\author[0000-0003-4335-0900]{Andrea Banzatti}
\affiliation{Department of Physics, Texas State University, 749 North Comanche Street, San Marcos, TX 78666, USA}
\email{banzatti@txstate.edu}

\author[0000-0003-3401-1704]{Naman S. Bajaj}
\affiliation{Lunar and Planetary Laboratory, The University of Arizona, Tucson, AZ 85721, USA}
\email{namanbajaj@arizona.edu}

\author[0000-0001-7962-1683]{Ilaria Pascucci}
\affiliation{Lunar and Planetary Laboratory, The University of Arizona, Tucson, AZ 85721, USA}
\email{pascucci@arizona.edu}

\author[0000-0002-1575-680X]{James Miley}
\affiliation{Departamento de F\'isica, Universidad de Santiago de Chile, Avenida Victor Jara 3659, Santiago, Chile}
\affiliation{Millennium Nucleus on Young Exoplanets and their Moons (YEMS), Chile}
\affiliation{Center for Interdisciplinary Research in Astrophysics and Space Exploration (CIRAS),
Universidad de Santiago, Santiago, Chile}
\email{james.miley@alma.cl}

\author[0000-0003-0787-1610]{Geoffrey A. Blake}
\affiliation{Division of Geological \& Planetary Sciences, MC 150-21, California Institute of Technology, Pasadena, CA 91125, USA}
\email{gab@caltech.edu}

\author[0000-0003-3682-6632]{Colette Salyk}
\affiliation{Vassar College, 124 Raymond Avenue, Poughkeepsie, NY 12604, USA}
\email{cosalyk@vassar.edu}

\author[0000-0003-2251-0602]{John M. Carpenter}
\affiliation{Joint Atacama Large Millimeter/submillimeter Array Observatory, Alonso de C\'ordova 3107, Vitacura, Santiago, Chile}
\email{John.Carpenter@alma.cl}

\author[0000-0001-8764-1780]{Paola Pinilla}
\affiliation{Mullard Space Science Laboratory, University College London, Holmbury St Mary, Dorking, Surrey RH5 6NT, UK}
\email{p.pinilla@ucl.ac.uk}

\author[0000-0002-2828-1153]{Lucas A. Cieza}
\affiliation{Instituto de Estudios Astrof\'isicos, Universidad Diego Portales, Avenida Ejercito 441, Santiago, Chile}
\email{lucas.cieza@mail.udp.cl}

\author[0000-0002-4147-3846]{Miguel Vioque}
\affiliation{European Southern Observatory, Karl-Schwarzschild-Stra{\ss}e 2, 85748 Garching bei M\"{u}nchen, Germany
}
\email{miguel.vioque@eso.org}

\author[0000-0002-1103-3225]{Beno\^it Tabone}
\affiliation{Institut d'Astrophysique Spatiale, Universit\'e Paris-Saclay, CNRS, B\^atiment 121, 91405 Orsay Cedex, France}
\email{benoit.tabone@universite-paris-saclay.fr}

\begin{abstract}

We present JWST/MIRI Medium Resolution Spectrometer spectra of the wide-separation (projected separation $=980\,\mathrm{au}$) binary protoplanetary disks Sz~65 (K7; $0.68\,M_{\odot}$) and Sz~66 (M3; $0.30\,M_{\odot}$), reduced using the uniform pipeline of the JWST Disk Infrared Spectral Chemistry Survey. Both disks show rich molecular emission, including H$_2$O, CO$_2$, HCN, C$_2$H$_2$, and OH. The scaled spectra of the two disks exhibit remarkably similar H$_2$O, CO$_2$, and HCN line emission in the 13--18~$\mu$m region, with the only notable difference being stronger C$_2$H$_2$ emission in the primary (Sz~65). Beyond 18~$\mu$m, the difference in H$_2$O line emission between the two disks increases. Both the flux ratios and the slab-model-derived mass ratios of cold to hot H$_2$O ($\sim200$ to $\sim750\,\mathrm{K}$) and warm to hot H$_2$O ($\sim450$ to $\sim750\,\mathrm{K}$) are significantly higher in the secondary (Sz~66). Because binary stars share nearly the same age and metallicity, and as both disks appear compact in millimeter emission ($<30\,\mathrm{au}$), we suggest that the excess cold H$_2$O in the secondary is best explained by its unstructured dust disk, in contrast to the primary, which shows gaps at 6 and $20\,\mathrm{au}$. The enhanced cold water in the secondary is consistent with efficient pebble drift across the water snow line and increased H$_2$O vapor from the sublimation of icy mantles. Our results demonstrate that wide-separation binaries can serve as powerful control samples for isolating the impact of individual disk properties on inner-disk chemistry and evolution.

\end{abstract}


\section{Introduction} \label{sec:intro}

Planets form in protoplanetary disks and therefore their final composition strongly depends on the prevailing chemical composition of their natal disks. The inner disk ($<10\,\mathrm{au}$) is believed to be the region where terrestrial planets and the cores of gas giants assemble \citep{Williams11}. Therefore, during the first few million years of disk evolution, the chemical composition of nascent planetary bodies is largely imprinted by the physical and chemical processes in the inner disk \citep{Oberg21, Drazkowska23}. 

Mid-IR spectra have been widely applied to probe the inner regions of protoplanetary disks \citep{Carr08, Pontoppidan10, Salyk11, Pascucci13}. The Infrared Spectrometer (IRS) on the Spitzer Space Telescope (Spitzer) observed $\sim300$ protoplanetary disks with a spectral resolution of $R\sim600$ and detected various molecular species: CO is detected in most of the disks, whereas H$_2$O, CO$_2$, OH, HCN, and C$_2$H$_2$ are detected in nearly half of the T Tauri star disks \citep{Pontoppidan10}. However, the spectral lines obtained from Spitzer are unresolved and often suffer from blending. Recent JWST/MIRI observations achieve a much higher spectral resolution of $R\approx2000\text{--}3500$ \citep{Pontoppidan24}, revealing a rich molecular inventory in protoplanetary disks. This includes low-abundance isotopologues such as $^{13}$CO$_2$, $^{13}$CCH$_2$, H$^{13}$CN, C$^{18}$O$^{16}$O, and H$_2^{18}$O \citep{Grant23, Colmenares24, Salyk25, Salyk26} and more complex organic molecules such as $\mathrm{C}_{4}\mathrm{H}_{2}$ and C$_6$H$_6$ (benzene; \citealt{Tabone23, Arabhavi24}). In addition, the increased spectral resolution has enabled detailed modeling of molecular temperatures and column densities.

Recent JWST/MIRI surveys---notably the JWST Disk Infrared Spectral Chemistry Survey (JDISCS; \citealt{Pontoppidan24, Arulanantham25}) and the Mid-Infrared Disk Survey (\citealt{Henning24})---have revealed a remarkable diversity in protoplanetary disk chemistry. Observations range from water-rich spectra  \citep{Xie23, Grant24, Arulanantham25} to those dominated by CO$_2$ \citep{Grant23, Vlasblom25} or enriched in complex hydrocarbons \citep{Tabone23, Arabhavi24, Arabhavi25}, the latter predominantly found in disks around very low-mass hosts.

In addition, significant contrasts in H$_2$O temperature distributions have been found \citep{Banzatti23b, Gasman23, RomeroMirza24}. Recent studies have proposed that the diversity of H$_2$O line fluxes can be shaped by disk size and substructures \citep{Banzatti23b, Krijt25}. However, \cite{Temmink25} have found a wide diversity in H$_2$O abundances within a sample of relatively compact ($<60$\,au) disks, and the H$_2$O line flux may also depend on evolutionary state \citep{Banzatti17} and metallicity \citep{Guadarrama22}, which complicates isolating the effect of disk structure. 

Binary disks share nearly identical stellar ages, initial metallicities, and environments, therefore offering unique opportunities to investigate the relationship between mid-IR spectra and disk properties. Previous mid-IR studies have focused on relatively close ($<300$\,au) pairs of binaries \citep{Grant24, Kurtovic25}, where dynamical interactions alter the evolution of both disks in the binary \citep{Zagaria23}. This can sometimes result in either no detectable disk or only very weak molecular emission from the secondary. In contrast, wide pairs ($>300$\,au) mostly host unperturbed disks, making their evolution comparable to that of single stars. This is supported by millimeter emission surveys, which show that the disk luminosity distribution of wide pairs is similar to that of single stars, while close pairs are significantly fainter \citep{Harris12, Akeson19, Manara19, Panic21}.

Sz~65 and Sz~66 are a wide-separation (projected distance $\approx1000$\,au) binary at a distance of approximately 154\,pc identified using Gaia EDR3 data \citep{Bohn22}. The stellar and disk parameters are summarized in Table \ref{tab:stellarparam}. Based on stellar evolutionary tracks, the stellar masses are estimated to be $0.68^{+0.20}_{-0.14}\,M_{\odot}$ and $0.30^{+0.05}_{-0.04}\,M_{\odot}$ for Sz~65 and Sz~66, respectively \citep{Pascucci16, Deng25}.

The two stars are separated by an angular distance of $6.36^{\prime\prime}$ ($980\,\mathrm{au}$ on the sky plane), and therefore both disks are unlikely to be tidally truncated unless the orbits are highly eccentric \citep{Miley24}. \cite{Bohn22} statistically determined that the two stars have a high probability of being gravitationally associated based on Gaia proper motion measurements. Although \cite{Majidi23} suggests they may not be currently gravitationally bound based on radial velocity measurements, it is generally assumed that they formed at approximately the same time within similar environments. Therefore, comparing Sz~65 and Sz~66 with JWST/MIRI spectroscopy can illuminate the evolution of coeval systems with different stellar masses and luminosities.

The dust continuum and the CO emission of the binary have been characterized in detail with multiple Atacama Large Millimeter/submillimeter Array (ALMA) observations, making them compelling JWST targets for probing the corresponding warm inner-disk chemistry.
\cite{Miley24} study the dust continuum of the binary at high spatial resolution, and the ALMA Survey of Gas Evolution of Protoplanetary Disks (AGE-PRO) survey \citep{Deng25, Zhang25, Trapman25} studies gas disks at high spectral resolution. The disk size is characterized by $R_{90}$, the radius within which 90\% of the flux is contained. For the primary (Sz~65), the radius of the dust disk is $27\,\mathrm{au}$, with a shallow gap confirmed at $20\,\mathrm{au}$ and a tentative gap at $6\,\mathrm{au}$; the secondary (Sz~66) has a compact dust disk of $16\,\mathrm{au}$ with no substructure resolved at the same resolution ($\sim3\,\mathrm{au}$; \citealt{Miley24}). AGE-PRO measurements provide gas and dust mass estimates as shown in Table \ref{tab:stellarparam}. 

{\renewcommand{\arraystretch}{1.1}
\begin{deluxetable}{cccc}[t]
\tabletypesize{\scriptsize}
\tablecaption{Sz~65 and Sz~66 Star and Disk Parameters \label{tab:stellarparam}}
\tablehead{
  \colhead{Parameter} & \colhead{Sz~65} & \colhead{Sz~66} & \colhead{Reference}
}
\startdata
Distance (pc) & $153.0 \pm 0.6$ & $154.4 \pm 0.6$ & \cite{Bailer21} \\
Class & II             & II             & \cite{Alcala14} \\
Spectral Type & K7             & M3             & \cite{Alcala17} \\
$M_*$ ($M_{\odot}$) & $0.68^{+0.20}_{-0.14}$ & $0.30^{+0.05}_{-0.04}$ & \cite{Deng25} \\
$L_*$ ($L_{\odot}$) & $0.87$ & $0.22$ & \cite{Deng25} \\
$T_\mathrm{eff}$ (K) & $4060 \pm 187$ & $3415 \pm 79$ & \cite{Alcala17} \\
$R_*$ ($R_{\odot}$) & $1.84 \pm 0.40$ & $1.29 \pm 0.30$ & \cite{Alcala17} \\
$\log_{10} \dot{M}_\mathrm{acc}$ ($M_{\odot}$yr$^{-1}$) & $<-9.48$ & $-8.51$ & \cite{Manara23} \\
$M_\mathrm{dust}$ ($M_{\oplus}$) & $19.4$ & $3.0$ & \cite{Deng25} \\
$\log_{10} M_\mathrm{gas}$ ($M_{\odot}$) & $-2.72^{+0.20}_{-0.17}$ & $-3.12^{+0.40}_{-0.33}$ & \cite{Trapman25} \\
$R_\mathrm{90, dust}$ (au) & $27 \pm 4$ & $16 \pm 4$ & \cite{Miley24} \\
$R_\mathrm{90, CO}$ (au) & $142 \pm 15$ & $48 \pm 15$ & \cite{Miley24} \\
Inclination ($^\circ$) & $63 \pm 1$ & $68 \pm 1$ & \cite{Miley24} \\
\enddata
\end{deluxetable}
}

We organize this paper as follows. Section \ref{sec:obs} briefly describes the observations of Sz~65 and Sz~66 and the JDISCS data reduction procedure. Section \ref{sec:analysis} presents the results of line-ratio diagnostics and slab-model fitting. Section \ref{sec:discussion} compares the disk properties of the binary and discusses their implications for pebble drift theories. Section \ref{sec:summary} provides a summary and conclusions.

\section{Observations and Data Reduction} \label{sec:obs}

\subsection{Observations}\label{sec:observe}

The JWST/MIRI data used in this work are available at MAST at DOI: \dataset[10.17909/nnc1-9g02]{\doi{10.17909/nnc1-9g02}}.
Sz~65 and Sz~66 were observed with JWST/MIRI in Medium Resolution Spectrometer (MRS) mode on 2024 July 11 (UT) as part of the Cycle 2 GO program 3034 (PI: K. Zhang). The two targets were observed within the same visit, with a total scheduled visit duration of 3.55 hr. The MIRI/MRS spectra were obtained using the full MRS wavelength coverage to produce continuous spectra across 4.9--28\,$\mu $m. Observations used the standard four-point negative dither pattern optimized for point sources.

Data reduction for both disks follows the standard JDISCS procedure described in \cite{Pontoppidan24}, which adopts the standard MRS pipeline \citep{Bushouse25} up to Stage 2b. An asteroid reference observation of 515 Athalia was obtained using the same instrumental configuration and dither pattern. The asteroid calibration provides high-quality fringing removal and characterizes the relative spectral response function, maximizing the signal-to-noise ratio (S/N) in Channels 2--4. We use JDISCS reduction version 9.1, which includes improved bad-pixel correction and an updated long-wavelength ($\gtrsim20\,\mu$m) flux calibration.


\subsection{Continuum Subtraction}\label{sec:contsub}

We follow the continuum subtraction technique described in \cite{Pontoppidan24}. The continuum is obtained by iteratively applying median filters, retaining the lower flux density components of the spectra, and subsequently smoothing with a second-order Savitzky--Golay filter \citep{Savitzky64}. Broad molecular emission regions including 13.4--14.1\,$\mu $m (C$_2$H$_2$ and HCN) and 14.9--15.0\,$\mu $m (CO$_2$) are excluded from the continuum estimation. Following the approach in \citet{Banzatti25}, we identify a set of line-free regions and compare them to the continuum, then apply a final wavelength-dependent flux offset to the continuum.

\begin{figure}[htbp]
\centering
\includegraphics[width=0.47\textwidth]{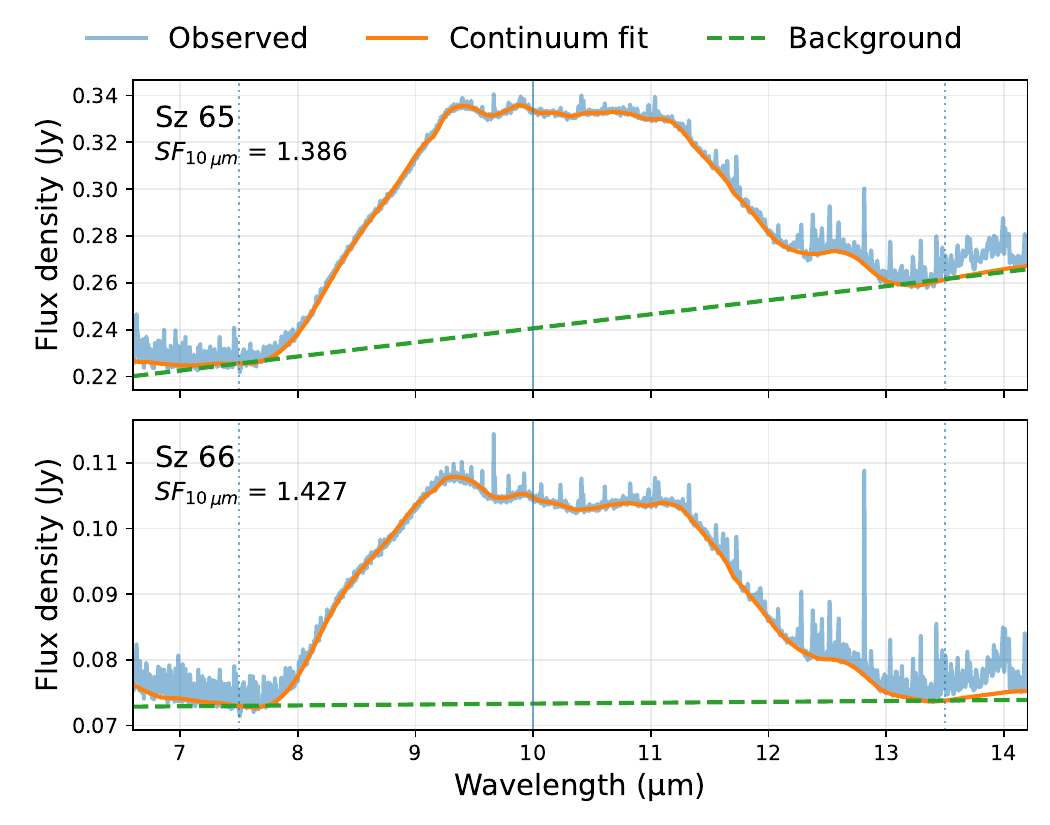}
\caption{The 10\,$\mu $m silicate feature of the primary Sz~65 (upper), and the secondary Sz~66 (lower). The green dashed line shows the interpolated background from 7.5 to 13.5\,$\mu $m.
\label{fig:silicate}}
\end{figure}

\begin{figure*}[htbp]
\includegraphics[width=0.98\textwidth]{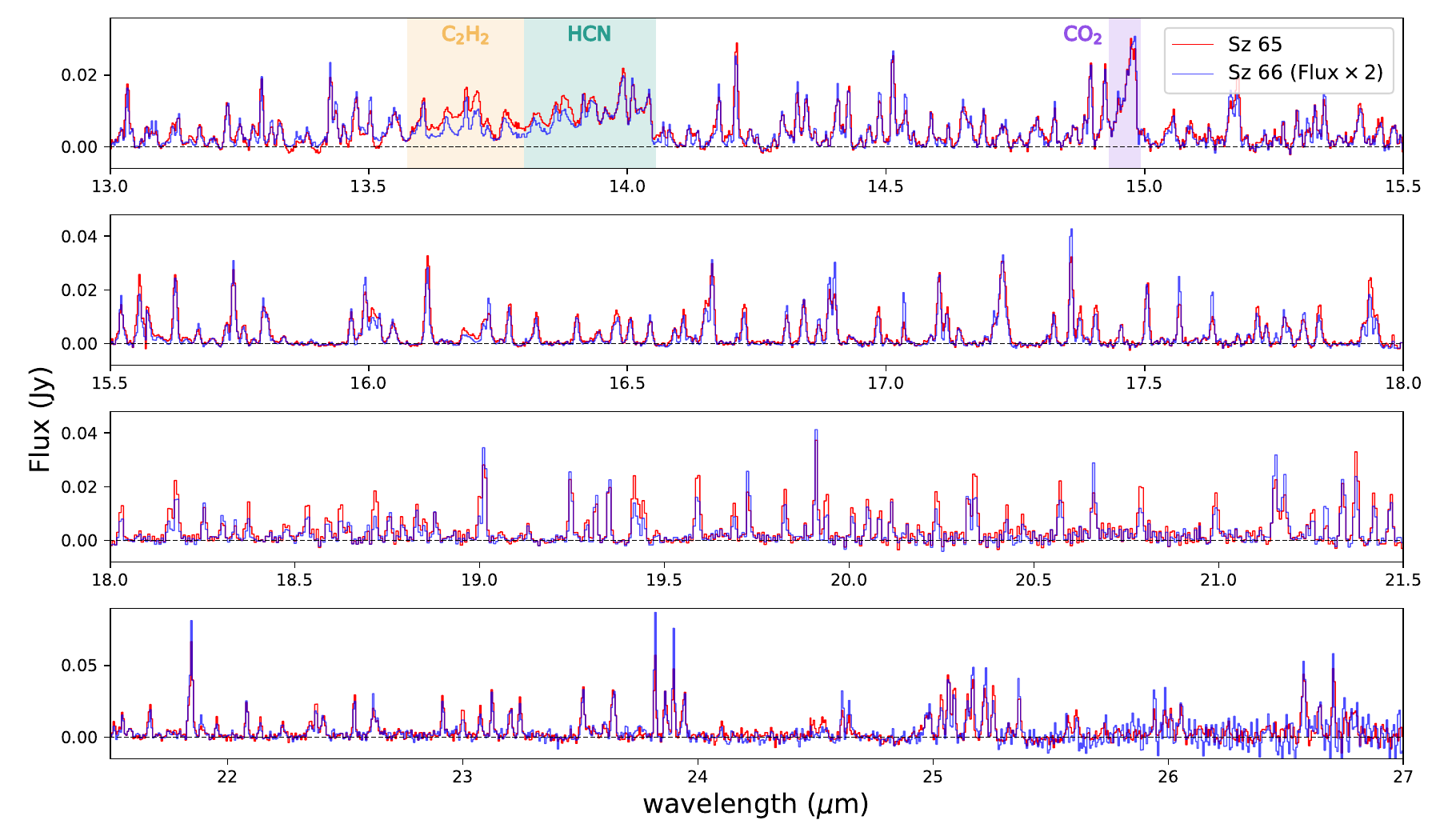}
\caption{The continuum-subtracted spectra of the primary Sz~65 (red), and the secondary Sz~66 (blue). The flux of the secondary is scaled by a factor of 2 to approximately match the lines at 13--18\,$\mu $m. Main emission features of C$_2$H$_2$, HCN, and CO$_2$ are highlighted.
\label{fig:Overlap}}
\end{figure*}

The continuum estimation is well determined for the primary across the full MIRI wavelength range of 4.9--28\,$\mu $m, although noise increases beyond 27\,$\mu $m. The secondary exhibits absorption signatures between 4.9--5.5\,$\mu $m, which degrade the quality of the continuum subtraction at the short-wavelength end. The absorption mainly comes from the M-type star's photosphere (see Appendix \ref{App:Photo}), and the disk atmosphere might contribute some absorption at high disk inclination (i = 68$^\circ$), as seen in \cite{Zhang15}. Therefore, as we will discuss in Section \ref{sec:4.9}, we are only able to model the CO P-branch emission ($\approx5\,\mu $m) for the primary. Figure \ref{fig:sz65cont} and Figure \ref{fig:sz66cont} in Appendix \ref{App:A} show the full MIRI spectra and the continuum estimation process; overall, the primary’s dust continuum is about 3 times brighter than the secondary’s, consistent with the contrast in stellar luminosity.

After obtaining the estimated continuum, we quantify the strength of the 10\,$\mu $m silicate feature by calculating $SF_{10} \equiv {F_{10}}/{F_\mathrm{bg}}$, where $F_{10}$ is the silicate continuum at 10\,$\mu $m and $F_\mathrm{bg}$ is obtained by linearly interpolating between 7.5 and 13.5\,$\mu $m; the $SF_{10}$ of Class II disks typically ranges from 1.2 to 3.5 \citep{KesslerSilacci06}. Figure \ref{fig:silicate} shows that the two disks have similar silicate feature strengths ($SF_{10}=1.39$ for Sz~65 and $SF_{10}=1.43$ for Sz~66). A reduced silicate feature is generally associated with grain growth and vertical settling \citep{Dullemond04, Lommen10}. Therefore, the nearly equal $SF_{10}$ values suggest broadly similar levels of dust grain processing in the binary disks.

\section{Analysis} \label{sec:analysis}

The continuum-subtracted spectra of Sz~65 and Sz~66 in the 13--27\,$\mu $m range are shown in Figure \ref{fig:Overlap}. After scaling up the flux of Sz~66 (the secondary) by a factor of 2, the emission of CO$_2$ (14.97\,$\mu $m) and HCN ($\sim14\,\mu $m) are very similar, while the C$_2$H$_2$ emission (13.6--13.8\,$\mu $m) is substantially stronger in Sz~65. The H$_2$O emission lines of the two disks show comparable intensities between 13 and 18\,$\mu $m. Beyond 18\,$\mu $m, where H$_2$O lines have generally lower upper-level energies ($E_\mathrm{up}$), the scaled spectrum of the secondary shows an excess in H$_2$O emission. In the following sections, we further examine the disk properties of the binary based on the observed emission features.

\subsection{Water Line-ratio Diagnostics} \label{sec:diag}

\begin{figure*}[htbp]
\centering
\includegraphics[width=0.9\textwidth]{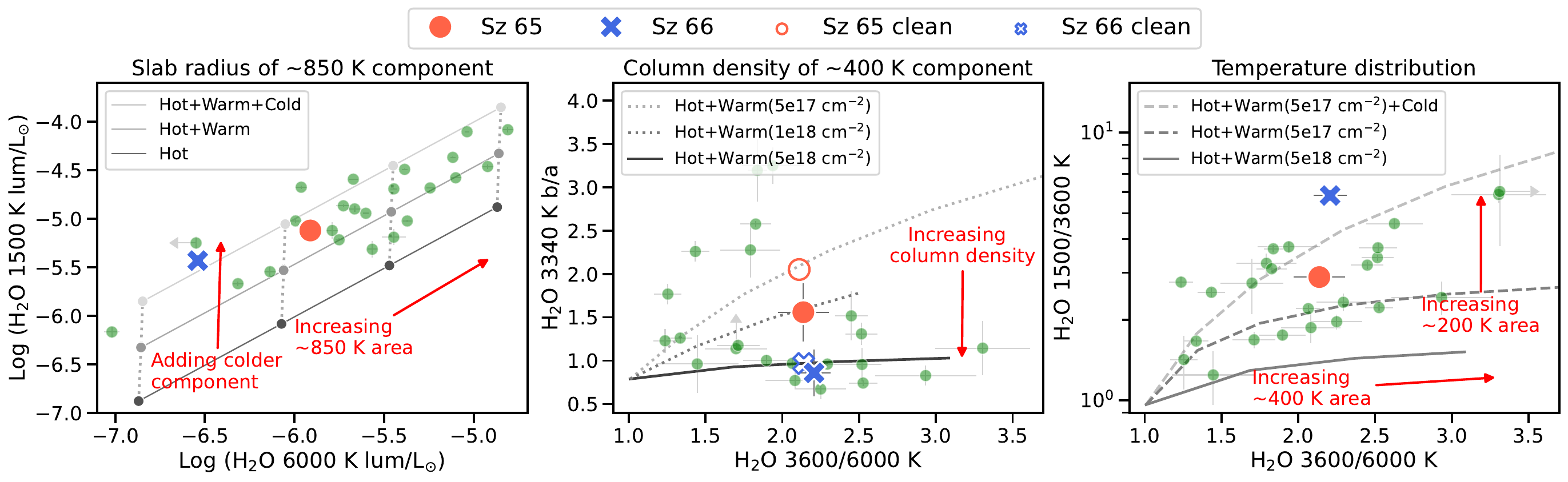}
\caption{Line diagnostic diagrams for Sz~65 and Sz~66, together with reference disks from \cite{Banzatti25} (green) and slab-model predictions (gray). Values for reference disks are updated to JDISCS version 9.0 (see Section 2.2 of \cite{Mallaney26}). \textit{Left}: The radius of the hot (850 K) component increases from left to right with the 6000 K line luminosity, and the warm and cold components increase from bottom to top with the 1500 K line luminosity. \textit{Middle}: The warm component increases from left to right, and the column density decreases from bottom to top. Hollow symbols indicate line ratios after removing organic contamination. \textit{Right}: The warm component increases from left to right, and the cold component increases from bottom to top. See Figure 10 in \cite{Banzatti25} for related details.
\label{fig:diag}}
\end{figure*}

\begin{figure*}[htbp]
\centering
\includegraphics[width=0.9\textwidth]
{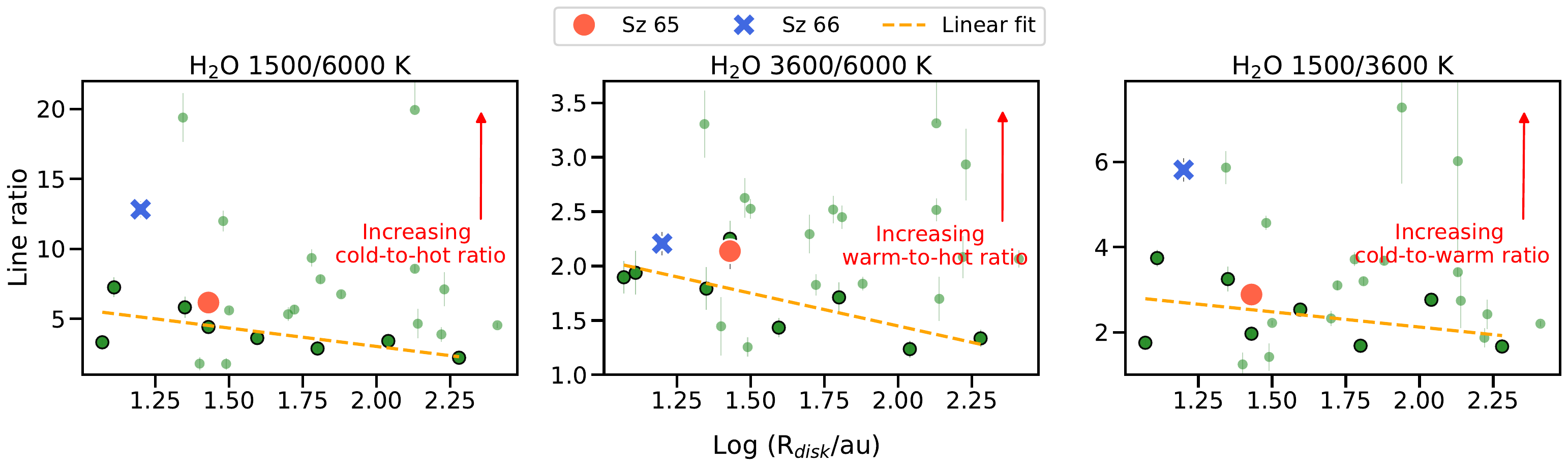}
\caption{Line diagnostic diagrams showing line ratios versus the dust disk size $R_{90}$. The three line pairs represent the flux ratio of cold/hot (left), warm/hot (middle), and cold/warm (right) H$_2$O. The anticorrelation (orange) is fitted from a subset of single stars without cloud contamination or a disk cavity (black outline).
See Figure 13 in \cite{Banzatti25} for details.
\label{fig:Rdisk}}
\end{figure*}

\begin{figure}[htbp]
\centering
\includegraphics[width=0.358\textwidth]{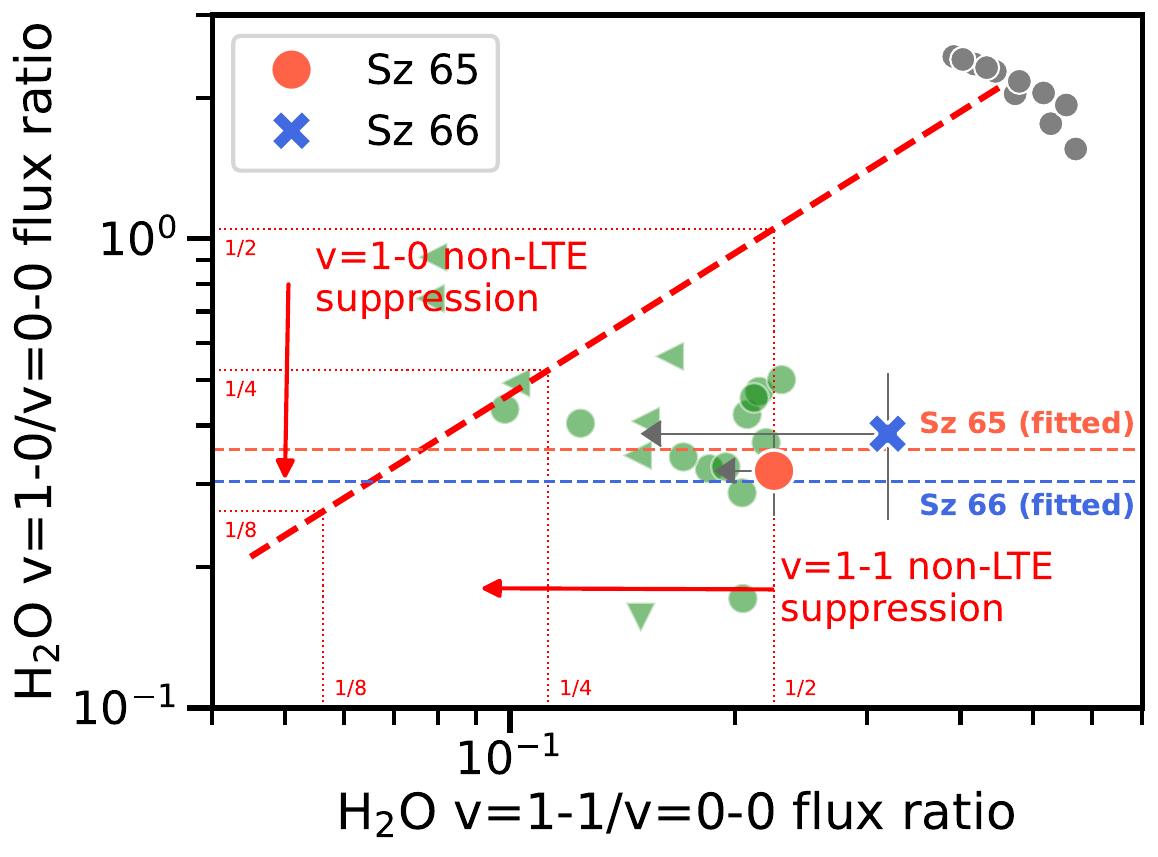}
\caption{Line diagnostic diagram for non-LTE suppression. Triangles and gray arrows indicate upper limits of undetected non-LTE lines for reference disks and the studied binary. The gray dots show LTE-predicted ratios, and the red dashed line denotes the level of suppression. Horizontal dashed lines indicate the best-fit suppression factors obtained in \ref{sec:5.5}. See Figure 7 in \cite{Banzatti25} for details.
\label{fig:nonLTE}}
\end{figure}

\begin{figure}[htbp]
\centering
\includegraphics[width=0.358\textwidth]{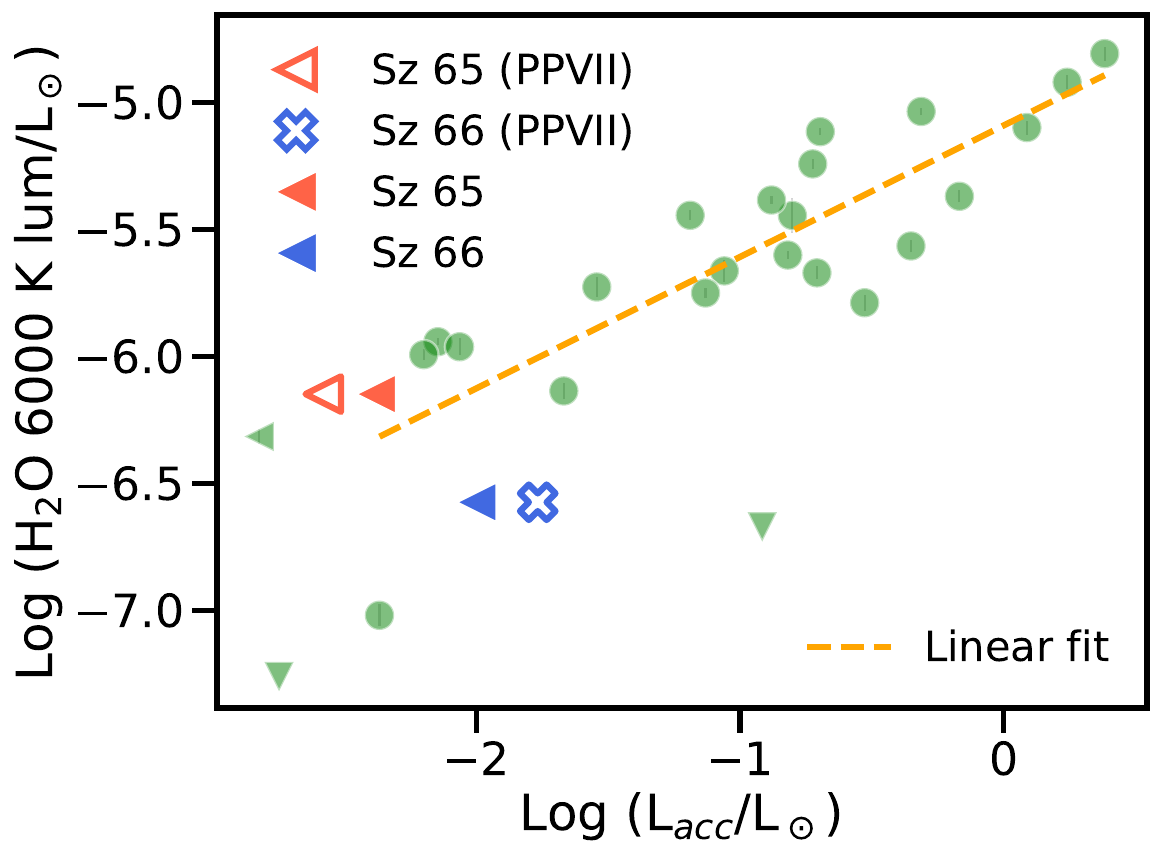}
\caption{Line diagnostics diagram showing accretion luminosity derived from mid-IR \ion{H}{1} lines versus the 6000 K line luminosity. Triangles indicate upper limits. For Sz~65 and Sz~66, the accretion luminosities reported in \cite{Manara23} (PPVII) derived from Balmer-jump fitting are shown with hollow marks. The correlation (orange) is fitted from reference disks with upper limits excluded.
\label{fig:Lacc}}
\end{figure}

We analyze the H$_2$O line fluxes of Sz~65 and Sz~66 using line-ratio diagnostics recently developed by \cite{Banzatti25}. These diagnostics provide a fast preliminary estimate to compare the temperature distribution, column density, emitting area, and departures from local thermodynamic equilibrium (LTE) of H$_2$O within the JDISCS samples, using a selected set of relatively isolated H$_2$O lines that span a range of upper-level energies ($E_\mathrm{up}$) and Einstein $A$ coefficients ($A_\mathrm{ul}$).

We measure line fluxes using \texttt{iSLAT} \citep{Jellison24, Johnson24} and use the line-ratio diagnostics to provide a preliminary estimate of the hot ($\sim850$\,K), warm ($\sim400$\,K), and cold ($\sim190$\,K) H$_2$O components of the binary. The lines involved in the diagnostics are labeled by their upper-level energies ($E_\mathrm{up}$), and details of the transitions are listed in Table 1 of \cite{Banzatti25}. Figure \ref{fig:diag} displays the line luminosities and ratios for Sz~65 and Sz~66 together with 25 reference Class II disks from \cite{Banzatti25}.  Slab models with temperatures of 190, 400, and 800\,K and column densities of $5\times10^{17}$, $1\times10^{18}$, and $5\times10^{18}\,\mathrm{cm}^{-2}$ are overlaid to provide estimates of disk properties. In the left panel, the vertical distance from the pure hot H$_2$O baseline indicates the amount of warm and cold H$_2$O components. The horizontal spread in the left panel is dominated by the slab radius of hot H$_2$O, suggesting that the primary (Sz~65) has a larger emitting area than the secondary (Sz~66). The middle panel uses the flux ratio (b/a ratio) of two lines with different $A_\mathrm{ul}$ and the same $E_\mathrm{up}$ to estimate the column density. Note that the transition at 13.50312\,$\mu $m used to calculate the b/a ratio might suffer from contamination by HCN and C$_2$H$_2$ emissions, causing an overestimation of the column density. Therefore, we recalculate the b/a ratios after subtracting organic emissions modeled in Section \ref{sec:models}, which are labeled as ``clean'' line ratios in the middle panel. The warm component in the primary shows a column density of $5\times10^{17}\,\mathrm{cm^{-2}}$ with subtraction and $1\times10^{18}\,\mathrm{cm^{-2}}$ without subtraction, whereas the secondary is close to $5\times10^{18}\,\mathrm{cm^{-2}}$. In the right panel, the cold H$_2$O component increases from bottom to top, demonstrating that the secondary has a more dominant cold water component.

Figure \ref{fig:Rdisk} overlays the binary on a diagnostic diagram of line ratio versus the size of the dust disk. The more compact Sz~66 has higher cold/hot ($E_\mathrm{up}$ = 1500 K/6000 K), warm/hot ($E_\mathrm{up}$ = 3600 K/6000 K), and cold/warm line ratios as compared to Sz~65.

Non-LTE effects near the disk surface suppress the v = 1--0 and v = 1--1 H$_2$O lines relative to v = 0--0 because of the difference in critical densities \citep{Banzatti23a}, which affect most of the lines from 5.5 to 8.5\,$\mu $m. Figure \ref{fig:nonLTE} shows the line ratios between three selected lines with vibration level v = 1--0 (8.0696\,$\mu $m), v = 1--1 (24.91403\,$\mu $m), and v = 0--0 (16.27136\,$\mu $m). The v = 1--0 to v = 0--0 ratios for Sz~65 and Sz~66 are suppressed by a factor of $\sim6$ compared to LTE models, consistent with the bulk distribution of other disks. The v = 1--1 line is not detected in either disk.

Figure \ref{fig:Lacc} places the binary within the correlation between hot H$_2$O emission and accretion luminosity ($L_\mathrm{acc}$) shown by previous studies \citep{Salyk11, Banzatti20, Banzatti25}. The $L_\mathrm{acc}$ is derived from mid-IR \ion{H}{1} lines \citep{Tofflemire25}; we also add $L_\mathrm{acc}$ of the binary reported in \cite{Manara23} (PPVII) derived from Balmer-jump fitting, which is generally considered more reliable \citep{Alcala14}. The \ion{H}{1}-derived $L_\mathrm{acc}$ is an upper limit for both disks; the Balmer-jump-derived $L_\mathrm{acc}$ is an upper limit for the primary, but reports a detection for the secondary slightly higher than the \ion{H}{1}-derived upper limit. According to Balmer-jump fitting, the secondary's $L_\mathrm{acc}$ is at least an order of magnitude higher than the primary, which does not align with the correlation between H$_2$O line flux and $L_\mathrm{acc}$.

In conclusion, the water line-ratio diagnostics provide a quick look at the binary’s water reservoirs, indicating a larger hot water-emitting area for the primary, larger warm and cold water contributions and a higher warm H$_2$O column density for the secondary, and comparable v = 1--0 non-LTE suppression.

\subsection{Slab Models of Molecular Emission}\label{sec:models}

The continuum-subtracted MIRI spectra reveal a forest of molecular lines. We detect H$_2$O lines throughout the full MIRI range for both disks, appearing primarily as pure rotational lines at longer wavelengths (13--27\,$\mu $m) and as rovibrational lines at shorter wavelengths (5.5--8.5\,$\mu $m). We detect significant emission from CO$_2$ (15.0\,$\mu $m), C$_2$H$_2$ ($\sim13.7\,\mu $m), HCN ($\sim14.0\,\mu $m), and CO (4.9--5.15\,$\mu $m). The isotopologue emission of $^{13}$CO$_2$ is analyzed in Appendix \ref{App:iso}.

We retrieve the physical parameters of molecular species by fitting the MIRI spectrum with LTE slab models, employing a Markov Chain Monte Carlo (MCMC) approach. Although we cannot spatially resolve inner-disk emission, the rich set of molecular lines with different $E_\mathrm{up}$ and $A_\mathrm{ul}$ allows us to probe the molecular reservoir with a wide range of temperatures and opacities. The line transition information is obtained from the HITRAN database \citep{Gordon22}. In previous studies, the slab model is typically characterized by three parameters: temperature ($T$), column density ($\log N$), and emitting area ($\log A$) \citep{Grant23,Pontoppidan24}. In this study, we replace $\log N$ and $\log A$ with $\log N-\log A$ and $\log N+\log A$, respectively. The latter term captures the observable mass ($\log M$) divided by the molecular mass, which remains constrained in the optically thin regime. This choice stabilizes the MCMC run by avoiding multiple degenerate parameters.

{\renewcommand{\arraystretch}{1.3}
\newcommand{\best}[1]{\begingroup\boldmath\textbf{$#1$}\endgroup}

\begin{deluxetable}{ccccc}[t]

\tabletypesize{\scriptsize}
\tablecaption{Best-fit model parameters of Sz~65 and Sz~66\label{tab:model6566}}
\tablehead{
  \colhead{Species} & \colhead{$T$ (K)} &
  \colhead{$\log N$ (cm$^{-2}$)} &
  \colhead{$\log M$ ($M_{\oplus}$)} &
  \colhead{$\log A$ (au$^{2}$)}
}
\startdata
\noalign{\vskip 1.5pt}
\multicolumn{5}{c}{\textbf{Sz~65}}\\
\noalign{\vskip 1.5pt}\hline
H$_2$O (Hot)               & \best{772^{+20}_{-9}} & $18.5^{+0.05}_{-0.05}$ & $-6.72^{+0.04}_{-0.07}$ & \best{-1.31^{+0.02}_{-0.04}}\\
H$_2$O (Warm)              & \best{442^{+15}_{-8}} & $18.3^{+0.07}_{-0.06}$ & $-6.26^{+0.07}_{-0.07}$ & \best{-0.63^{+0.03}_{-0.03}}\\
H$_2$O (Cold)              & \best{218^{+2}_{-12}} & $15.2^{+1.06}_{-0.22}$ & $-6.10^{+0.16}_{-0.04}$ & \best{2.62^{+0.37}_{-0.86}}\\
HCN                        & \best{837^{+46}_{-30}} & $17.6^{+0.07}_{-0.13}$ & \best{-8.06^{+0.03}_{-0.08}} & \best{-1.91^{+0.07}_{-0.04}}\\
C$_2$H$_2$                 & \best{1\,398^{+2}_{-59}} & $15.8^{+0.29}_{-3.72}$ & \best{-8.77^{+0.02}_{-0.03}} & \best{-0.74^{+3.70}_{-0.30}}\\
CO$_2$                     & $314^{+29}_{-20}$ & \best{17.9^{+0.18}_{-0.09}} & \best{-6.76^{+0.21}_{-0.20}} & \best{-1.08^{+0.06}_{-0.13}}\\
OH\tablenotemark{a} (Cold) & $444$ & $17.4$ & $-6.80$ & $-0.28$\\
OH\tablenotemark{a} (Warm) & $1877$ & $14.8$ & $-9.33$ & $-0.20$\\
OH\tablenotemark{a} (Hot)  & $5197$ & $13.1$ & $-11.91$ & $-1.08$\\
\noalign{\vskip 2pt}
\hline
$^{13}$CO$_2$\tablenotemark{b} & 258 & 13.2 & -8.46 & 1.89\\
CO\tablenotemark{c} & $1\,989^{+104}_{-2}$ & 17.0\tablenotemark{c} & $-8.09^{+0.02}_{-0.06}$ & -1.29\tablenotemark{c}\\
\hline\hline
\noalign{\vskip 1.5pt}
\multicolumn{5}{c}{\textbf{Sz~66}}\\
\noalign{\vskip 1.5pt}\hline
H$_2$O (Hot)               & $697^{+13}_{-18}$ & \best{19.0^{+0.10}_{-0.04}} & \best{-6.70^{+0.10}_{-0.06}} & $-1.73^{+0.03}_{-0.04}$\\
H$_2$O (Warm)              & $364^{+8}_{-13}$ & \best{18.7^{+0.07}_{-0.07}} & \best{-6.00^{+0.10}_{-0.07}} & $-0.75^{+0.06}_{-0.02}$\\
H$_2$O (Cold)              & $194^{+2}_{-8}$   & \best{15.6^{+0.60}_{-0.40}} & \best{-5.87^{+0.12}_{-0.04}} & $2.50^{+0.50}_{-0.46}$\\
HCN                        & $702^{+53}_{-35}$ & \best{17.7^{+0.14}_{-0.10}} & $-8.24^{+0.08}_{-0.10}$ & $-2.12^{+0.05}_{-0.07}$\\
C$_2$H$_2$                 & $1\,342^{+58}_{-30}$ & \best{16.6^{+0.92}_{-1.83}} & $-9.34^{+0.04}_{-0.04}$ & $-2.15^{+1.80}_{-0.85}$\\
CO$_2$                     & \best{355^{+25}_{-40}} & $17.5^{+0.16}_{-0.20}$ & $-7.46^{+0.22}_{-0.22}$ & $-1.42^{+0.15}_{-0.04}$\\
OH\tablenotemark{a} (Cold) & $284$ & $17.4$ & $-6.39$ & $0.19$\\
OH\tablenotemark{a} (Warm) & $1084$ & $14.0$ & $-9.21$ & $0.80$\\
OH\tablenotemark{a} (Hot)  & $3164$ & $13.8$ & $-10.28$ & $-0.14$\\
\noalign{\vskip 2pt}
\hline
$^{13}$CO$_2$\tablenotemark{b} & 212 & 14.2 & -8.45 & 0.93\\
\enddata
\tablecomments{Boldface marks the numerically larger value between Sz~65 and Sz~66 for each species and parameter.}
\tablenotetext{a}{We include OH components to subtract overlapping OH emission from blended lines; the listed parameters are not representative of thermalized conditions, and no uncertainties are included.}
\tablenotetext{b}{The uncertainties cannot be derived with confidence due to limited residual signals, and the best-fit values are coarse estimates.}
\tablenotetext{c}{For CO in the optically thin regime, $\log N$ and $\log A$ are fully degenerate; we therefore do not report their uncertainties.}
\end{deluxetable}
}

We include the optical-depth effect of overlapping lines by adopting Gaussian line profiles with an FWHM of $\Delta v_\mathrm{opacity} = 4.7\,\mathrm{km\,s^{-1}}\;(\sigma = 2\,\mathrm{km\,s^{-1}})$, consistent with the values adopted by \cite{Salyk11} and \cite{Grant23}. The LTE model is implemented with \texttt{spectools-ir} \citep{Salyk22}, which has been proven successful in reproducing MIRI spectra \citep{Pontoppidan24, Arulanantham25} and is also benchmarked against IRS data \citep{Salyk11, Carr11}. The MCMC technique is deployed using \texttt{emcee} \citep{Foreman13} to retrieve the posterior distribution of physical parameters. In the following, we present the slab-model fitting results in decreasing order of wavelength.

\subsubsection{13--27\,$\mu $m: Main Molecular Models}\label{sec:13}

We divide the 13--27\,$\mu$m wavelength range into two intervals at 18\,$\mu$m because of the spectral resolution gap between MRS Channels 3 and 4 \citep[see Figure~6 of][]{Pontoppidan24}. The molecular species and the FWHMs of lines are treated separately in the two intervals, whereas the $\chi^2$ values are combined in the MCMC technique. We determine the FWHM due to instrument resolution and Keplerian motion by fitting selected H$_2$O lines. For 13--18\,$\mu $m, the FWHMs of Sz~65 and Sz~66 are $\Delta v_\mathrm{obs}$ = 150\,$\mathrm{km\,s^{-1}}$ and 121\,$\mathrm{km\,s^{-1}}$; for 18–27\,$\mu $m, the FWHM are $\Delta v_\mathrm{obs}$ = 169\,$\mathrm{km\,s^{-1}}$ and 147\,$\mathrm{km\,s^{-1}}$. The systemic radial velocities measured from H$_2$O lines are 8.4\,$\mathrm{km\,s^{-1}}$ and 5.7\,$\mathrm{km\,s^{-1}}$ for Sz~65 and Sz~66, respectively, after converting to the local standard of rest frame, which are close to the radial velocities fitted in AGE-PRO \cite{Trapman25} (Sz~65: 4.5\,$\mathrm{km\,s^{-1}}$, Sz~66: 4.5\,$\mathrm{km\,s^{-1}}$), considering the velocity uncertainty of MIRI \citep{Argyriou23}. We estimate the noise level of the continuum-subtracted spectra using a Gaussian process following the approach in \cite{RomeroMirza24}; the results are summarized in Appendix \ref{App:B}. For 13–18\,$\mu $m, the mean noise levels of Sz~65 and Sz~66 are 1.98 mJy and 1.23 mJy; for 18–27\,$\mu $m, the noise levels are 4.40 mJy and 2.93 mJy. We include three H$_2$O components to approximate the radial temperature gradient of the disk, with temperatures initialized from flat distributions spanning $\pm 50$\,K around 800\,K (hot), 400\,K (warm), and 200\,K (cold). The hot and warm components enter both wavelength intervals, whereas the cold component is restricted to 18--27\,$\mu $m. We also fit CO$_2$, C$_2$H$_2$, HCN, and OH simultaneously in the MCMC runs. We represent OH with three temperature components to approximate its non-LTE effects as in \cite{Banzatti25}, but the parameters retrieved under the LTE assumption are not representative of the thermal conditions.

\begin{figure*}[htbp]
\includegraphics[width=\textwidth]{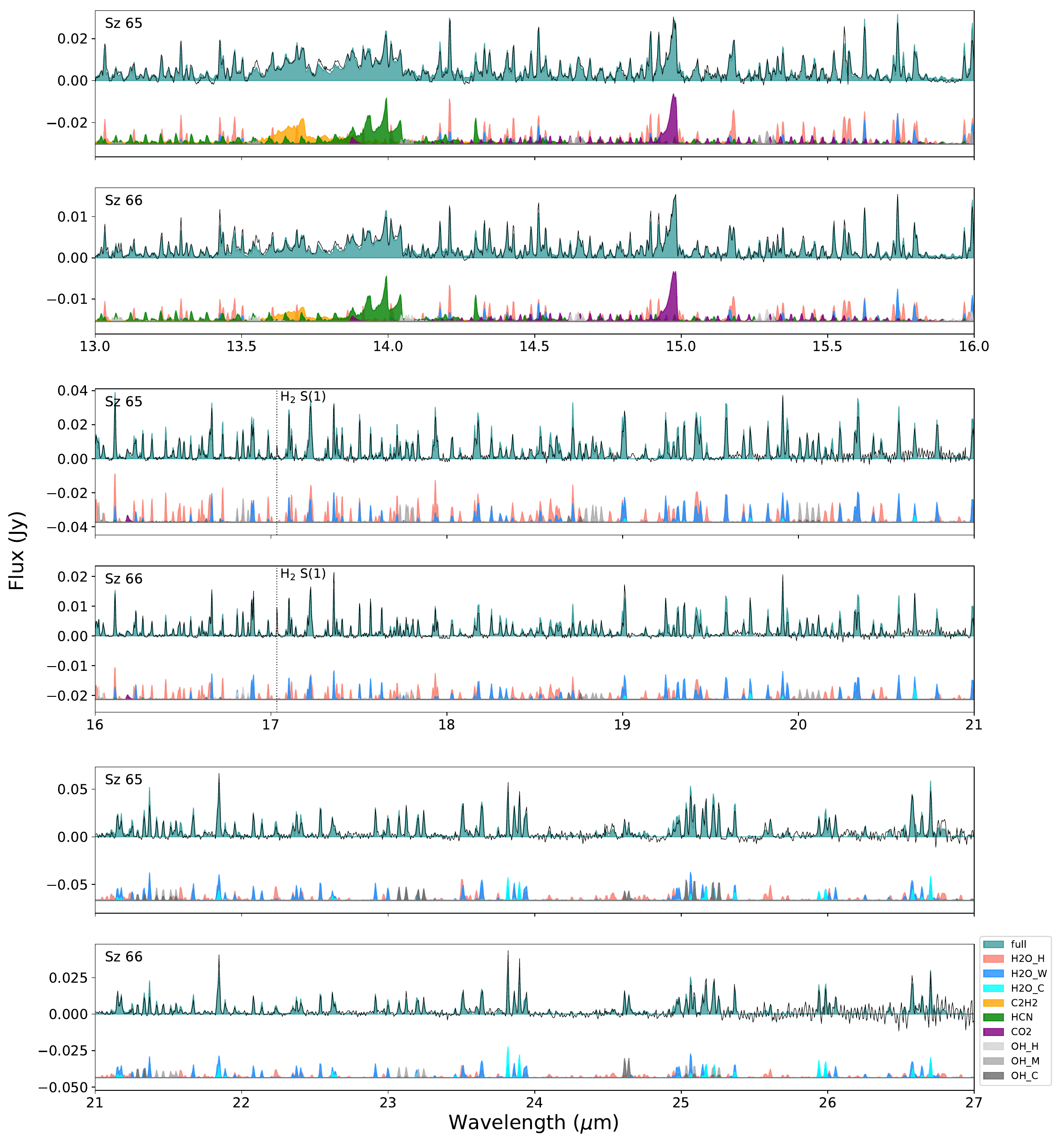}
\caption{The best-fit models and the continuum-subtracted spectra (black) of the primary Sz~65 and the secondary Sz~66 in the 13--27\,$\mu $m range. The total model flux is shown in cyan in the upper axis, while the molecular contributions are shown in the lower axis. The H$_2$ line, not included in the model, is indicated separately.
\label{fig:bestfit1}}
\end{figure*}

\begin{figure*}[!t]
\centering
\includegraphics[width=0.9\textwidth]{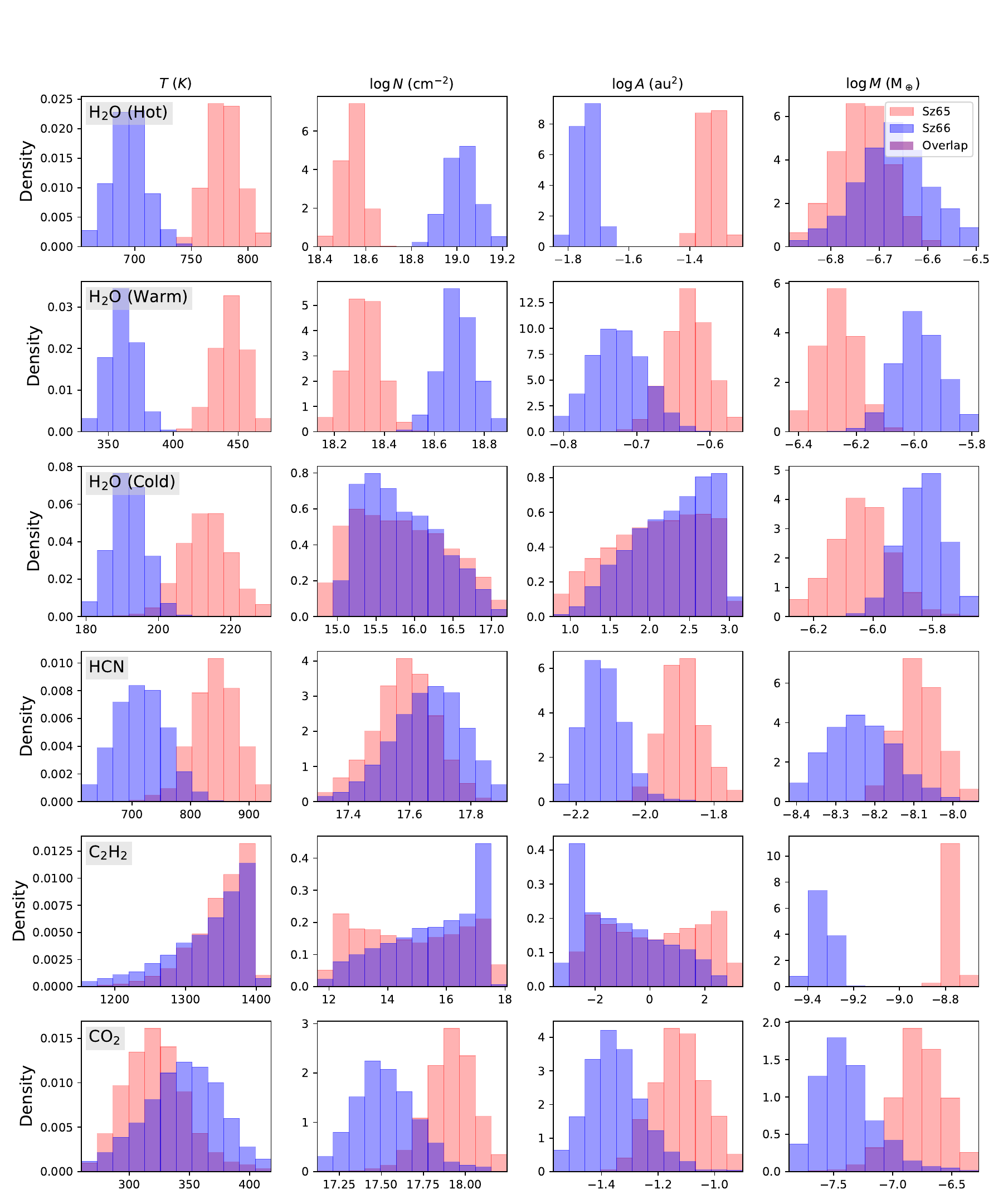}
\caption{The posterior distributions of the slab-model parameters of H$_2$O, CO$_2$, C$_2$H$_2$, and HCN for the primary Sz~65 (red) and the secondary Sz~66 (blue).
\label{fig:compareH2O}}
\end{figure*}

Figure \ref{fig:bestfit1} presents the best-fit models alongside the continuum-subtracted spectra for both disks. Table \ref{tab:model6566} summarizes the best-fit parameters and the 1$\sigma$ uncertainties. The larger values in the binary are represented in bold numbers. The emitting area ($\log A$) in Table \ref{tab:model6566} is calculated by dividing the observable mass by the column density and molecular mass. The uncertainties of the OH components are not included because their parameters are severely degenerate. Figure \ref{fig:compareH2O} compares the posterior distributions of the main model parameters.

\begin{figure*}[htbp]
\includegraphics[width=\textwidth]{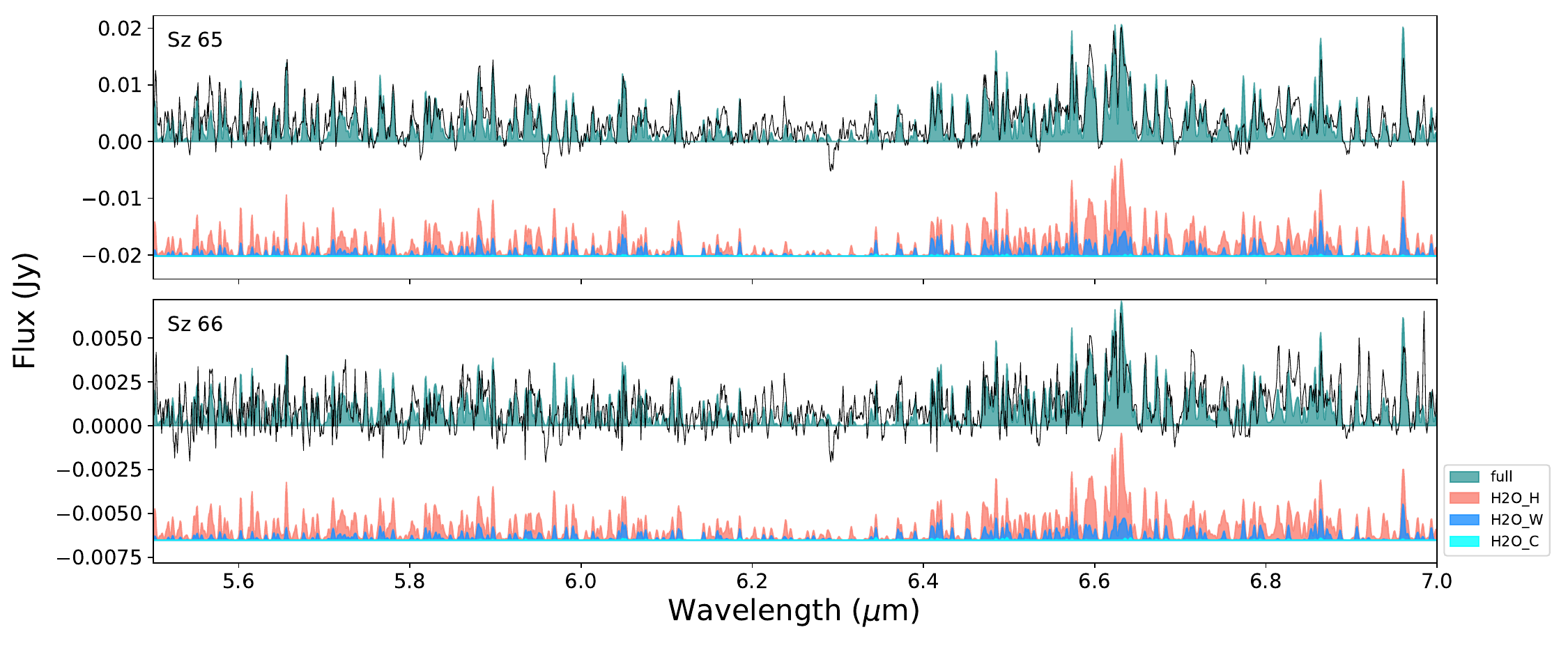}
\caption{The continuum-subtracted spectra and the model fluxes in 5.5--7\,$\mu$m scaled by the best-fit suppression factors obtained in Section \ref{sec:5.5}.
\label{fig:H2Oshort}}
\end{figure*}

Among the three H$_2$O components, the hot and warm components are well constrained for both disks, whereas the cold components exhibit degeneracy between column density and emitting area. The secondary (Sz~66) exhibits lower temperatures and smaller emitting areas across all components, consistent with its lower stellar luminosity. However, it possesses higher H$_2$O column densities and observable masses, particularly in the warm and cold components. 

The HCN and CO$_2$ components are also well constrained and show similar temperatures and column densities in the two disks, although the secondary has a slightly lower HCN temperature and CO$_2$ column density. The best-fit HCN and CO$_2$ masses are consistently higher in the primary. The C$_2$H$_2$ fits yield unconstrained column densities and unphysical temperatures pegged near the prior upper limit (1400\,K). 
The fitted temperature is well above the common value (800--1000\,K) seen in JDISCS samples \citep{Arulanantham25}. This issue is further discussed in Appendix \ref{App:underpred} together with the flux underprediction around 13.8\,$\mu $m.
Despite the unphysical best-fit parameters, Figure \ref{fig:bestfit1} clearly demonstrates that the C$_2$H$_2$ emission is substantially stronger in the primary, and Figure \ref{fig:compareH2O} indicates that the well-constrained C$_2$H$_2$ mass is higher in the primary.

\subsubsection{5.5--8.5\,$\mu $m: Non-LTE Water Emission}\label{sec:5.5}

We analyze the rovibrational H$_2$O emission in the 5.5--8.5\,$\mu $m range, where significant water emission exists and does not heavily overlap with CO emissions. It is known that rovibrational H$_2$O lines are often subthermally excited compared to the bulk pure rotational lines \citep{Meijerink09, Bosman22, Banzatti23a}. For each disk, we apply our three-component H$_2$O model derived from the 13--27\,$\mu $m fit and then scale the emission by a suppression factor. We note that non-LTE suppression may be temperature dependent \citep{Banzatti23a}; however, we assume a uniform suppression for simplicity within the scope of this work.

We obtained the best-fit suppression factors by minimizing $\chi^2$ for each disk, yielding 5.9 and 6.9 for Sz~65 and Sz~66. Figure \ref{fig:H2Oshort} shows that the scaled model fluxes reproduce the spectra of both disks from 5.5 to 7\,$\mu $m reasonably well, consistent with the result shown in Figure \ref{fig:nonLTE} using line-ratio diagnostics.

The observed non-LTE effects are attributed to low gas densities that are insufficient to thermalize the rovibrational emission \citep{Banzatti23a}. Although the secondary (Sz~66) has a smaller mass, the similar reduction factors indicate that it has a comparable gas density at the emitting layer of lines in the MIRI wavelength range, which may be related to the compactness of its disk.

\subsubsection{4.9--5.15\,$\mu $m: CO Models}\label{sec:4.9}

\begin{figure}[t]
\centering
\includegraphics[width=0.46\textwidth]{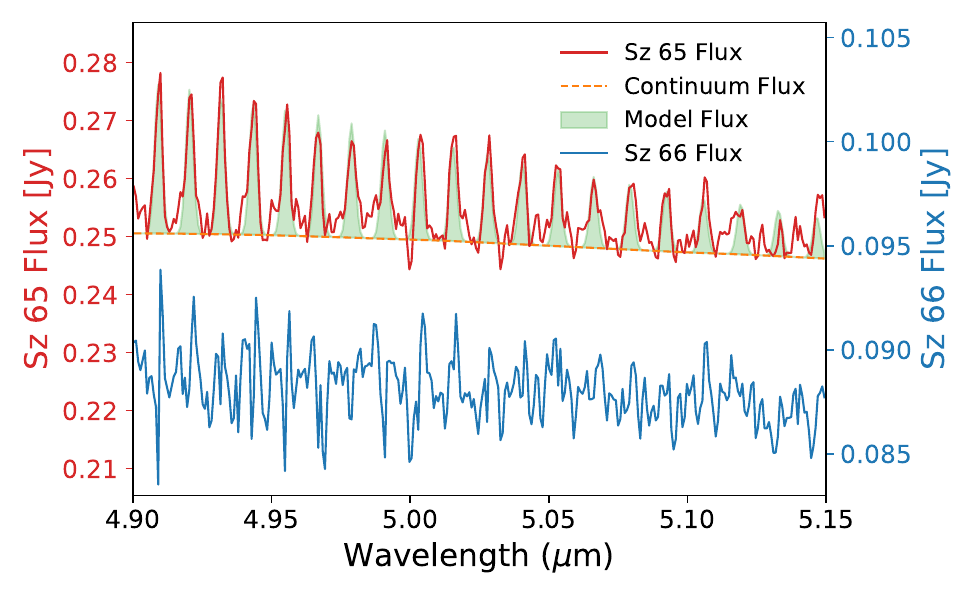}
\caption{The observed fluxes of the primary Sz~65 (red) and the secondary Sz~66 (blue) over 4.9--5.15\,$\mu $m, shown together with the continuum and CO model flux of the primary. The secondary is severely affected by photospheric absorption; therefore, the continuum estimation and model fitting are not performed.
\label{fig:CO}}
\end{figure}

MIRI captures the P branch of the fundamental CO rovibrational band near 5\,$\mu $m. When performing continuum subtraction, we obtain a relatively clean emission spectrum for the primary but not for the secondary due to photospheric absorption and possibly disk atmosphere absorption. Consequently, we model the CO emission only for the primary.

Rather than adopting the H$_2$O FWHM, we derive a separate FWHM from the CO line width, yielding $\Delta v_\mathrm{obs} = 190$\,$\mathrm{km\,s^{-1}}$. We select the 4.9--5.15\,$\mu $m region where CO emission is dominant over H$_2$O and subtract the overlapping H$_2$O flux modeled in Section \ref{sec:5.5}. Higher-vibrational CO lines are sensitive to non-LTE effects and are often weak or absent \citep{Banzatti22}. Therefore, we restrict the fit to the v = 1--0 lines.

{\renewcommand{\arraystretch}{1.1}
\begin{deluxetable*}{lcc ccc ccc}
\tablewidth{0pt}
\tablecaption{Atomic and H$_2$ emission of Sz~65 and Sz~66\label{tab:atomic}}
\tablehead{
\colhead{Line} & \colhead{$\lambda$} & \colhead{Band} & \multicolumn{3}{c}{Sz~65} & \multicolumn{3}{c}{Sz~66} \\[-5pt]
\colhead{} & \colhead{[$\mu$m]} & \colhead{} &
\colhead{FWHM} & \colhead{Doppler} & \colhead{Flux} &
\colhead{FWHM} & \colhead{Doppler} & \colhead{Flux} \\[-5pt]
\colhead{} & \colhead{} & \colhead{} &
\colhead{(km\,s$^{-1}$)} & \colhead{(km\,s$^{-1}$)} & \colhead{($10^{-16}$ erg s$^{-1}$ cm$^{-2}$)} &
\colhead{(km\,s$^{-1}$)} & \colhead{(km\,s$^{-1}$)} & \colhead{($10^{-16}$ erg s$^{-1}$ cm$^{-2}$)}
}
\startdata
\mbox{[\ion{Ne}{2}]} & 12.8135 & 3A & 126$\pm$5 & -26.5$\pm$2.1 & 32.8$\pm$1.7 & 99$\pm$1 & -22.8$\pm$0.5 & 28.5$\pm$0.5 \\
\mbox{[\ion{Ar}{2}]} & 6.9853 & 1C & 112$\pm$24 & -19.7$\pm$10.3 & 11.0$\pm$3.1 & 91$\pm$11 & -22.3$\pm$4.3 & 8.2$\pm$1.3 \\
H$_2$ S(1) & 17.0348 & 3C & 83$\pm$10 & -9.4$\pm$4.0 & 4.4$\pm$0.7 & 88$\pm$3 & -4.3$\pm$1.1 & 5.6$\pm$0.2 \\
H$_2$ S(3) & 9.6649 & 2B & 79$\pm$11 & -11.3$\pm$4.3 & 8.3$\pm$1.5 & 79$\pm$9 & -8.4$\pm$3.7 & 8.9$\pm$1.4 \\
\noalign{\vskip 2pt}
\hline
\noalign{\vskip 2pt}
\ion{H}{1} (6--5) & 7.4599 & 1C & 186$\pm$28\tablenotemark{a} & -89.8$\pm$12.1\tablenotemark{a} & 22.6$\pm$4.5\tablenotemark{a} & \nodata & \nodata & \nodata \\
\ion{H}{1} (7--6) & 12.3719 & 3A & 208$\pm$37\tablenotemark{a} & 13.2$\pm$15.8\tablenotemark{a} & 15.0$\pm$3.5\tablenotemark{a} & 160$\pm$31\tablenotemark{a} & 51.0$\pm$13.1\tablenotemark{a} & 3.2$\pm$0.8\tablenotemark{a} \\
H$_2$ S(2) & 12.2786 & 3A & 372$\pm$88\tablenotemark{a} & -122.9$\pm$37.1\tablenotemark{a} & 12.8$\pm$4.0\tablenotemark{a} & 192$\pm$23\tablenotemark{a} & -38.2$\pm$9.8\tablenotemark{a} & 7.9$\pm$1.2\tablenotemark{a} \\
H$_2$ S(5) & 6.9095 & 1C & \nodata & \nodata & \nodata & 91$\pm$14\tablenotemark{a} & -14.7$\pm$5.6\tablenotemark{a} & 6.9$\pm$1.4\tablenotemark{a} \\
\enddata
\tablecomments{The Doppler velocities are corrected to the stellar rest frame; the FWHMs are not deconvolved for the instrumental resolution.}
\tablenotetext{a}{Values are very uncertain because of overlapping molecular lines; the actual uncertainty might be higher than the 1 $\sigma$ error of Gaussian fitting listed.}
\end{deluxetable*}
}

Figure \ref{fig:CO} shows the continuum, observed fluxes, and model fluxes for the primary, together with the observed fluxes for the secondary. The best-fit CO parameters are shown in Table \ref{tab:model6566}. Note that the CO model is optically thin and degenerate. In contrast with the model for 13--27\,$\mu$m, the uncertainties in the column density and emitting area cannot be derived with confidence, and these parameters should be regarded only as order-of-magnitude estimates. The best-fit temperature is slightly higher than typical broad-component CO temperatures (800--1500\,K) derived from ground-based observations \citep{Banzatti15, Banzatti22}. Note that MIRI only covers the high-$J$ CO lines ($J \gtrsim 25$) that higher-resolution ground-based observations show to be increasingly dominated by a hotter, inner component \citep{Banzatti22}. Fluxes of lower-$J$ lines are needed to better constrain both the temperature and the optical depth of the emission; higher-resolution ground-based spectroscopy is also required to resolve the different excitation components that are blended in the MIRI spectra \citep{Banzatti23a,Wheeler24,Temmink24}.

\subsection{Atomic Lines and H$_2$ Lines}\label{sec:atomic}

Mid-IR atomic lines (e.g., [\ion{Ne}{2}] and [\ion{Ar}{2}]) and H$_2$ rotational lines are widely used to search for disk winds and jets (see \cite{Pascucci23} for a review). High-resolution spectroscopy has shown that [\ion{Ne}{2}] profiles and spatial extension can reveal blueshifted wind components or jet emission \citep{Pascucci09,vanBoekel09}, and \cite{Pascucci25} has recently demonstrated that H$_2$ can trace disk winds. Motivated by this, we search for these emissions in both the aperture-integrated spectra and the spatially extended MIRI emissions of the binary.

\begin{figure}[htbp]
\centering
\includegraphics[width=0.34\textwidth]{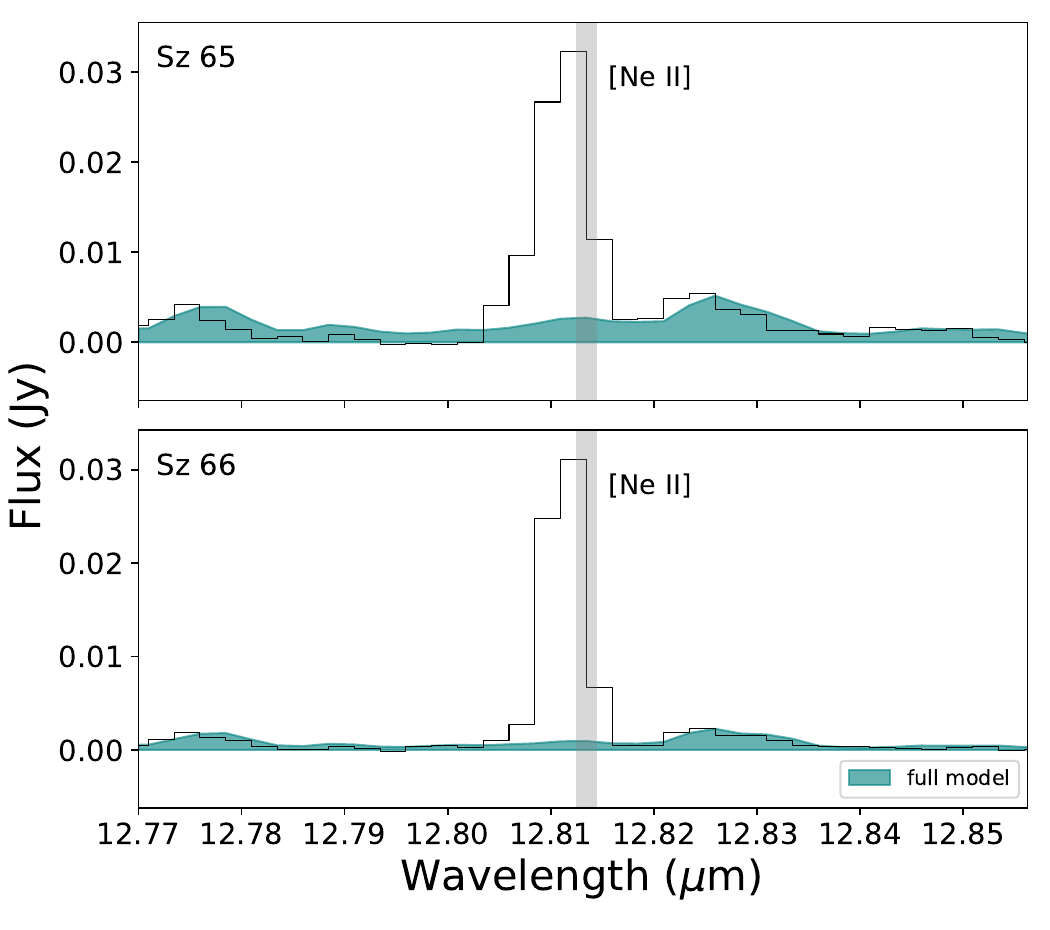}
\caption{[\ion{Ne}{2}] lines of Sz~65 and Sz~66. The rest wavelength is indicated by the gray line, and the cyan regions represent the molecular model fluxes obtained in Section~\ref{sec:13}.
\label{fig:NeII}}
\end{figure}

\begin{figure}[htbp]
\centering
\includegraphics[width=0.34\textwidth]{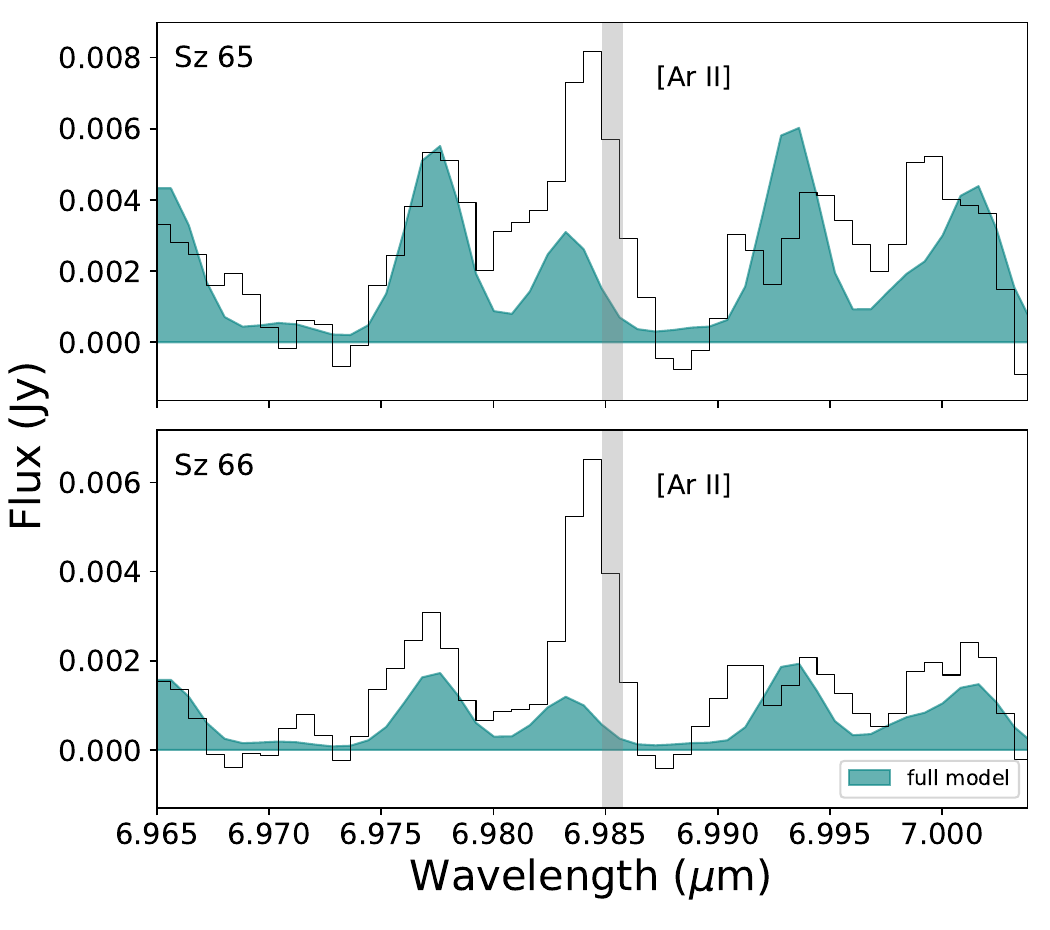}
\caption{[\ion{Ar}{2}] lines of Sz~65 and Sz~66; plotting conventions follow Figure \ref{fig:NeII}. The molecular models are scaled by the factors obtained in Section \ref{sec:5.5}.
\label{fig:ArII}}
\end{figure}

\begin{figure*}[htbp]
\centering
\includegraphics[width=1\textwidth]{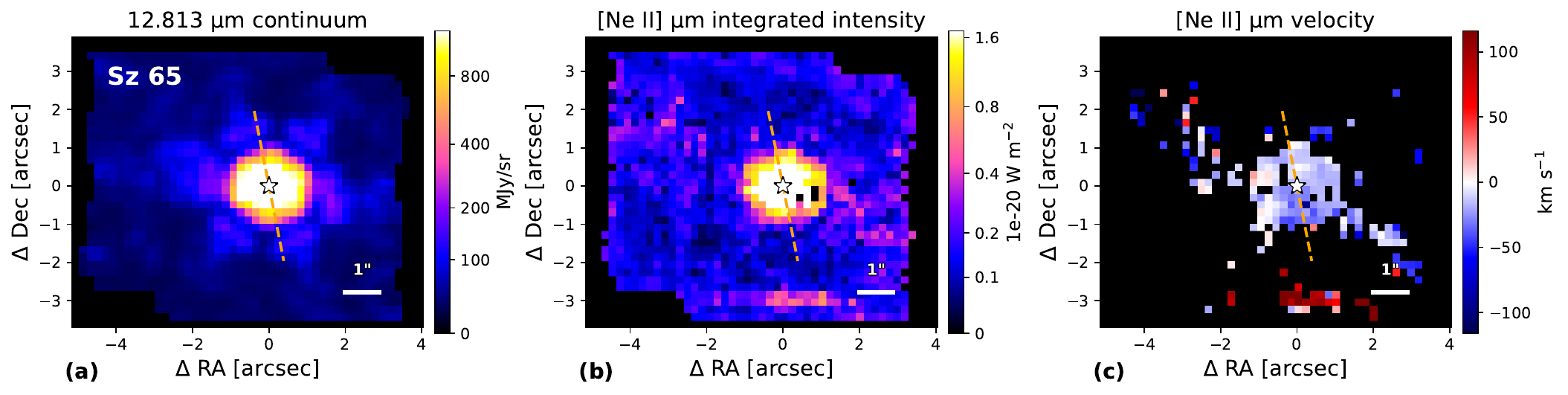}
\caption{(a) The line-subtracted continuum, (b) line intensity, and (c) Doppler velocity of the extended [\ion{Ne}{2}] emission of the primary (Sz~65). The orange dotted line marks the major axis of the disk.
\label{fig:Ne65}}
\end{figure*}

\begin{figure*}[htbp]
\centering
\includegraphics[width=1\textwidth]{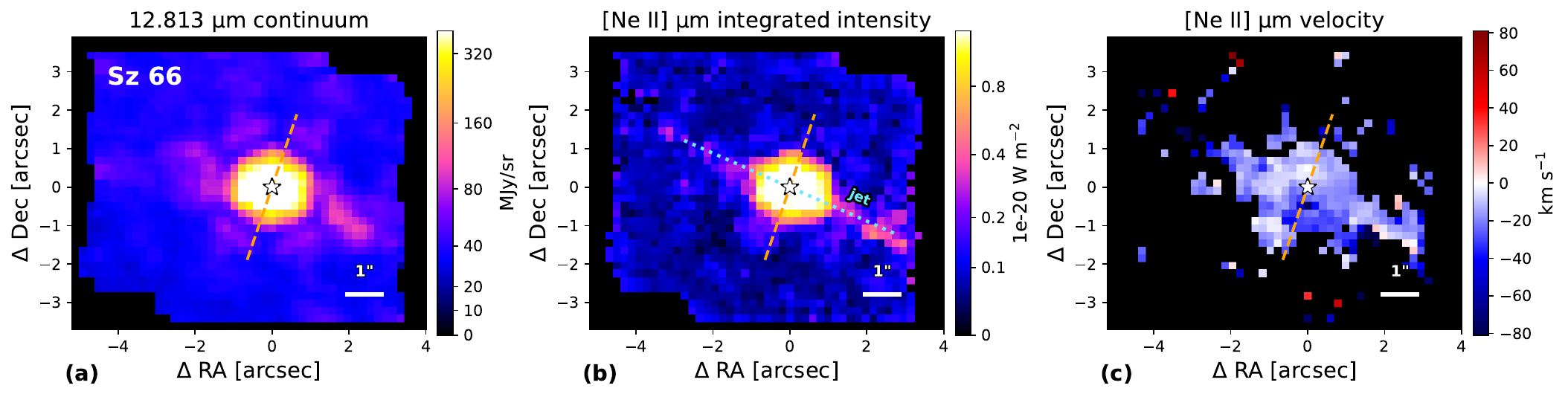}
\caption{(a) The line-subtracted continuum, (b) line intensity, and (c) Doppler velocity of the extended [\ion{Ne}{2}] emission of the secondary (Sz~66). A jet feature (cyan) perpendicular to the disk's major axis is tentatively detected.
\label{fig:Ne66}}
\end{figure*}

For the aperture-integrated spectrum, we subtract the molecular models in Section \ref{sec:models} to disentangle atomic emissions from a forest of molecular lines. We then measure the line fluxes and radial velocities using \texttt{iSLAT} \citep{Jellison24, Johnson24}. Table~\ref{tab:atomic} lists the Gaussian fit measurements for all detected atomic and H$_2$ lines in the aperture-integrated spectra. The Doppler velocities are corrected to the stellar rest frame. Uncertainties in molecular model subtraction degrade some line retrievals and also explain part of the nondetections. The atomic lines in the primary are generally broader and brighter. Figures~\ref{fig:NeII} and~\ref{fig:ArII} show the [\ion{Ne}{2}] and [\ion{Ar}{2}] profiles, which are the two best-constrained atomic lines; both are single peaked and blueshifted by $\sim20\,\mathrm{km\,s^{-1}}$, indicative of outflowing gas.

We retrieve the extended emission of both disks following the approach described in \cite{Bajaj24, Pascucci25, Bajaj25}. Significant extended emission of H$_2$ and [\ion{Ne}{2}] is detected in both disks. In particular, the [\ion{Ne}{2}] intensity map of the secondary (Sz~66) shows a tentative jetlike feature.

Figures \ref{fig:Ne65} and \ref{fig:Ne66} show the line-subtracted continuum, line intensity, and Doppler velocity of [\ion{Ne}{2}] extended emission of both disks. The Doppler velocity is corrected to the stellar rest frame. Orange dotted lines mark the direction of the major-axis position angle (PA) measured from ALMA continuum observations (PA$_{\mathrm{Sz65}} = 110^\circ$, PA$_{\mathrm{Sz66}} = 80^\circ$; \citealt{Miley24}). In the secondary, the intensity map traces a jet-shaped structure perpendicular to the major axis, whereas no structure can be confidently identified in the primary. We note that the [\ion{Ne}{2}] velocity map of the secondary does not reveal a clear blue--red dipole structure across the disk plane, which might be partially due to its high inclination (i = 68$^\circ$). Moreover, additional continuum emission is present in the same direction as the potential jet, which makes the jet detection of the secondary only tentative. Nevertheless, the overall blueshift observed toward both disks suggests that the [\ion{Ne}{2}] emission is likely tracing jets and/or atomic winds.

\section{Discussion} \label{sec:discussion}

\subsection{Binary Comparison}
Inner-disk chemical properties can be influenced by a range of factors, including stellar age, luminosity, metallicity, accretion rate, radiation environment, disk structure, and gravitational interactions. Consequently, isolating the impact of any single factor on disk chemistry is challenging. Wide-separation ($>300$\,au) binaries can serve as laboratories with controlled variables, sharing nearly identical stellar ages, metallicities, and environments. As a result, the remaining factors (e.g., stellar luminosity, accretion rate, and disk properties) become the principal variables.

Among the protoplanetary disk systems observed with JWST/MIRI, Sz~65 and Sz~66 constitute the first wide-separation Class~II binary analyzed in detail. Previous studies have analyzed the molecular lines of several Class II binaries with relatively small projected separations, including DF Tau ($9\,\mathrm{au}$) \citep{Grant24}, VW Cha ($130\,\mathrm{au}$), WX Cha ($143\,\mathrm{au}$), and RW Aur ($240\,\mathrm{au}$; \citealt{Kurtovic25}). The DF Tau system is not spatially or spectrally resolved by MIRI. The separations of VW Cha, WX Cha, and RW Aur are larger, but the secondary disks of all three systems are mostly line poor, which might be caused by disk truncation, low stellar mass, or an inner-disk cavity \citep{Kurtovic25}. In contrast, as a wide-separation binary, Sz~65 and Sz~66 show no evidence of dynamical perturbation and exhibit remarkable similarities in H$_2$O, CO$_2$, and HCN line emission. Therefore, we consider them ideal controlled samples for studying dynamically unperturbed disk evolution.

Comparisons of the continuum-subtracted spectra (Figure~\ref{fig:Overlap}) and the best-fit slab parameters (Table~\ref{tab:model6566}; Figure~\ref{fig:compareH2O}) reveal close similarities between the primary (Sz~65) and the secondary (Sz~66) in the CO$_2$, HCN, and hot ($\sim750\,\mathrm{K}$) H$_2$O components, with systematically scaled-down emitting areas in the secondary. This aligns with expectations for unperturbed binaries with the same metallicity, as the HCN/H$_2$O flux ratio has been proposed as a robust, temperature-insensitive indicator of the inner-disk C/O ratio \citep{Carr11, Najita13, Anderson21}. The C$_2$H$_2$/H$_2$O flux ratio is lower in the secondary; however, this C/O indicator might also reflect the luminosity contrast (0.87 and 0.22 $L_{\odot}$, respectively) according to \cite{Anderson21}.

The most notable difference is the cold water emission. Both the water diagnostic plots and the best-fit parameters indicate that the secondary has a significantly enhanced cold water mass relative to its hot water, HCN, and CO$_2$. The slab-model results of the secondary match the ``H$_2$O-dominated phase'' described in \cite{Sellek25}, where cold H$_2$O is enhanced while CO$_2$ remains largely unchanged.

Because the binary system controls for stellar age, metallicity, and the environment, we discuss three remaining factors that could drive the chemical differences. We examine stellar properties and accretion rate, two H$_2$O-regulating factors commonly discussed in the literature, and explore the impact of disk structure differences inferred from ALMA observations.

\begin{figure*}[htbp]
\centering
\includegraphics[width=0.95\textwidth]{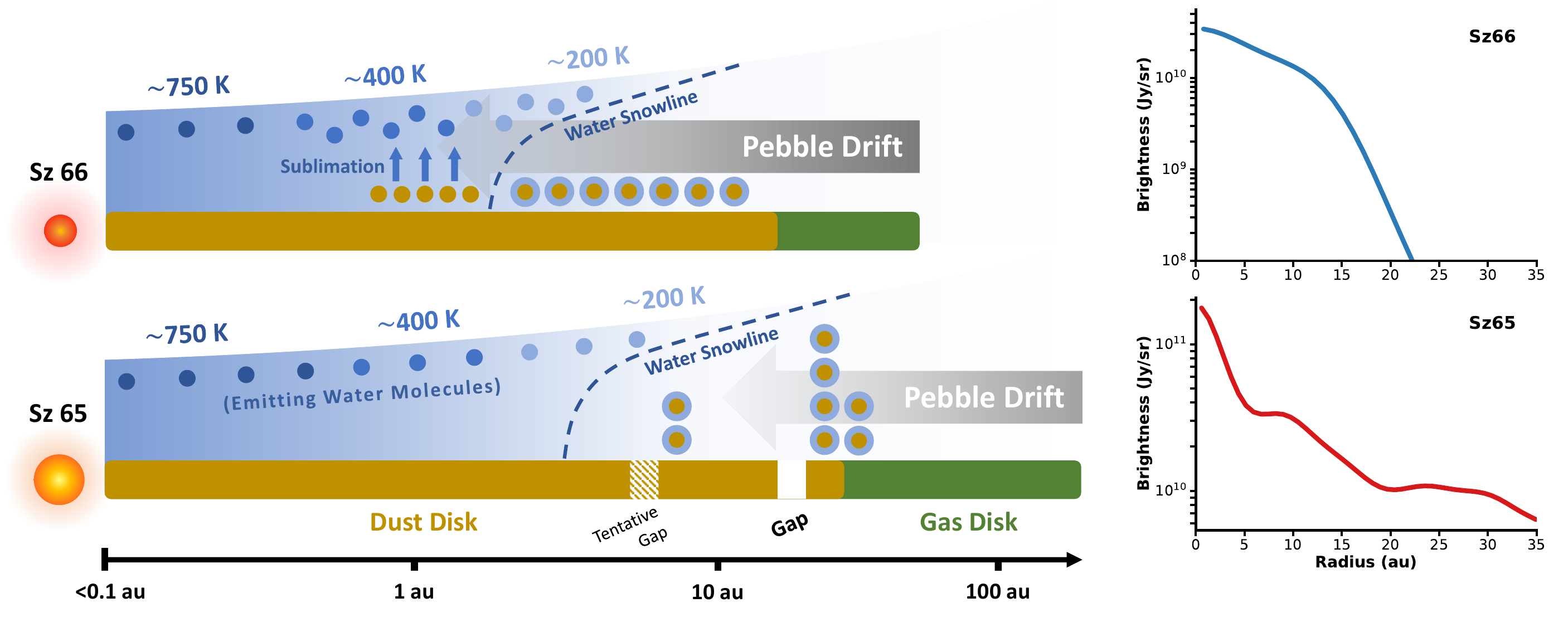}
\caption{Illustration of the disk structures and the pebble drift conditions of the two disks, with the millimeter-wavelength radial intensity profiles from \cite{Miley24} on the right. The intensity profiles are extracted from the uv plane of ALMA observations. Sz~66 has a smaller, unstructured disk that facilitates pebble drift, enriching the warm and cold water reservoirs through the sublimation of ice mantles, while the gaps of Sz~65 prevent pebbles from crossing the water snow line.
\label{fig:cartoon}}
\end{figure*}

\subsection{Explaining the Cold H$_2$O Excess}

\subsubsection{Stellar Properties}

The secondary (Sz~66) has a lower mass and stellar luminosity (0.30 versus 0.68\,M$_\odot$, and 0.22 versus 0.87\,L$_\odot$), which implies a closer snow line and a more compact water-emitting region. Under such a scenario, Sz~66 should exhibit lower water masses across all three components. Instead, we find the two sources have comparable masses of hot water and Sz~66 actually has significantly higher masses in warm and cold water than Sz~65, directly contradicting the expected trend.

In addition, a lower stellar mass may produce higher pebble drift velocities \citep{Pinilla13,Pinilla22}, though this effect is predicted to be significant primarily for very low-mass stars ($\lesssim0.2\,M_\odot$).


\subsubsection{Accretion Rate}

While the \ion{H}{1}-derived accretion luminosity ($L_\mathrm{acc}$) is an upper limit for both sources, the $L_\mathrm{acc}$ of the secondary (Sz~66) is at least an order of magnitude higher according to literature values \citep{Alcala17, Manara23}, as shown in Figure \ref{fig:Lacc}. Previous studies indicate that hot water line luminosity strongly correlates with $L_\mathrm{acc}$ \citep{Salyk11, Banzatti20, Banzatti25}; however, this correlation may partly reflect a shared evolutionary trend, as both $L_\mathrm{acc}$ and inner-disk H$_2$O emission weaken as disks disperse \citep{Pontoppidan10,Banzatti17}. Because the binary shares a common age, such evolutionary effects are controlled for, yet the primary still shows brighter hot water emission despite its lower $L_\mathrm{acc}$. This indicates that $L_\mathrm{acc}$ alone does not determine the hot water emission in this system. An additional possibility is that the secondary hosts a small inner dust cavity, which reduces hot water emission. For the warm and cold H$_2$O components, the correlation with $L_{\rm acc}$ is weaker because these lines originate at larger radii, where the connection to accretion heating is less direct \citep{Banzatti23b}. Therefore, the elevated warm and cold water masses in Sz~66 relative to Sz~65 are unlikely to be driven solely by differences in accretion.

Additionally, variable accretion has also been used to explain the cold H$_2$O enrichment in EX Lup \citep{Smith25} and XX Cha \citep{Temmink25}, in which an accretion outburst could liberate a large reservoir of cold H$_2$O. While this may remain a possible cause for the cold H$_2$O enrichment in Sz~66, we have not found any observational evidence that reports an accretion outburst in its history.

\subsubsection{Disk Structure}

Previous studies proposed that the sublimation of ice mantles on inward-drifting pebbles replenishes cold and warm water in the inner disk \citep{Ciesla06}. \cite{Kalyaan21} and \cite{Kalyaan23} relate the efficiency of pebble drift to the size of the dust disk and the presence of gaps: compact, featureless disks experience efficient inward drift and inner-disk water enrichment, while large, structured disks trap pebbles beyond the gap. The anticorrelation between cold water enrichment and dust disk size is observationally supported by Spitzer samples \citep{Banzatti20} and a relatively small JWST sample \citep{Banzatti23b}. In addition, \cite{Krijt25} found a correlation between inner-disk gap radius and cold water reservoir, suggesting that close-in gaps prevent cold water enrichment. \citet{Easterwood24} also suggest that close-in deep gaps and multiple gaps tend to reduce cold water delivery.

Figure \ref{fig:cartoon} illustrates the disk structure and the pebble drift conditions in the two disks. Also presented are the radial intensity profiles of the two dust disks from \cite{Miley24}, extracted from the uv plane of ALMA 1.3 mm observations using \texttt{FRANK} \citep{Jennings20} with $\alpha=1.2$ and $w_\mathrm{smooth}=10^{-4}$. The midplane H$_2$O snow line radii are roughly estimated to be at $3\,\mathrm{au}$ and $1.5\,\mathrm{au}$ for the primary and secondary, respectively, based on their stellar luminosities and standard midplane snow line scalings \citep{Mulders15}. The contrasting disk structures inferred from ALMA observations \citep{Miley24} are consistent with the theoretical expectations discussed. The primary Sz~65 has a larger dust disk ($27\,\mathrm{au}$) compared to the cold water-enriched secondary Sz~66 ($16\,\mathrm{au}$), following the anticorrelation between cold water enrichment and dust disk size as shown in Figure \ref{fig:Rdisk}. In addition, the primary has a shallow gap at $20\,\mathrm{au}$, and a tentative signal of an inner gap at $6\,\mathrm{au}$, while the secondary has a compact disk with no structure detected at $\sim3\,\mathrm{au}$ resolution. We note that the inner gap in Sz 65 might be deeper than observed due to resolution limits, as Sz 65 has a high inclination (63$^\circ$) and the gap width tends to be smaller at inner radii.

The spatial extents of the disk components provide further insight into the observed water discrepancies. The primary exhibits a large disk size ratio of 5.3:1 (gas at 142\,au versus dust at 27\,au), whereas the secondary has a smaller ratio of 3:1 (gas at 48\,au versus dust at 16\,au), although the latter ratio is more uncertain because the gas disk radius of Sz 66 is less well constrained. 
The large size ratio of the primary suggests that significant radial drift occurred in its outer-disk region \citep{Trapman_2020,Sanchis21}. However, its cold water emission is still weaker than the secondary's, implying that the inner-disk substructure plays a more important role. Studies have proposed that inner gaps can trap pebbles beyond the water snow line \citep{Kalyaan23, Easterwood24, Krijt25}. The inner (6\,au) gap of Sz 65 might correspond to the "traffic jam" regime modeled by \cite{Mah24}, which slows the H$_2$O supply into the inner disk. Conversely, the secondary’s featureless disk allows pebbles to drift efficiently across the snow line. Given the similarity between the binary disks, our study provides further evidence under controlled stellar age and metallicity that inner-disk gaps regulate pebble drift, consequently shaping the cold water reservoir in the inner disk.

According to AGE-PRO measurements, the millimeter dust masses of the primary and the secondary are 19.4 and 3.0 $M_{\oplus}$ \citep{Deng25}, and the gas masses are $1.9^{+1.1}_{-0.6}\times10^{-3}\,M_\odot$ and $7.6^{+11}_{-4.0}\times10^{-4}\,M_\odot$ \citep{Trapman25}. The larger contrast in millimeter dust mass is consistent with more efficient depletion of millimeter-sized grains in the secondary through radial drift and dust growth in a compact disk.

Efficient pebble drift may also increase the inner-disk opacity, partially obscuring molecular emission from the emitting layers \citep{Sellek25,Houge25}. Such an opacity effect could contribute to the overall lower molecular fluxes in the secondary, but it does not naturally explain the relative enrichment of the cold H$_2$O component.


Of the factors considered, differences in disk structure offer the most robust explanation for the observed features in the secondary: the excess of cold water, the lower dust-to-gas ratio, and the overall lower molecular fluxes.

\subsection{Uncertainties and Caveats}

We stress that analyzing a single pair of binary disks is insufficient to conclusively determine how dust disk structures shape inner-disk chemistry. Indeed, several uncertainties remain in this study: although we have argued that the accretion rate difference is unlikely to drive the cold water excess, its potential influence on longer-term disk evolution cannot be fully excluded with current data. In addition, two notable model underpredictions in the organic wavelength region remain unexplained (see Appendix \ref{App:underpred}), which might be related to the unphysical C$_2$H$_2$ model parameters. While a single pair of binaries could suffer from case-specific uncertainties, future observations on more wide-separation binary disks will greatly enhance our understanding of disk evolution under controlled age and metallicity.


\section{Summary and Conclusions}
\label{sec:summary}

We perform a detailed comparative analysis of the JWST/MIRI spectra of the wide-separation binary Sz~65 and Sz~66, primarily yielding the following results.

\begin{enumerate}
    \item The two disks exhibit remarkably similar inner-disk molecular emission. In particular, CO$_2$, HCN, and the hot ($\sim750\,\mathrm{K}$) H$_2$O component have comparable best-fit temperatures and column densities, with the secondary (Sz~66) showing systematically smaller emitting areas. The non-LTE suppression of H$_2$O lines in the 5.5--8.5\,$\mu $m range is also similar.
    \item Both the H$_2$O line-flux diagnostics and the LTE slab models indicate a significant excess of warm ($\sim450\,\mathrm{K}$) and cold ($\sim200\,\mathrm{K}$) water emission in Sz~66 relative to its hot H$_2$O, HCN, and CO$_2$. This is consistent with efficient pebble drift in its compact, unstructured disk.
    \item After subtracting the molecular slab models, we detect atomic emission in both disks, including [\ion{Ne}{2}] and [\ion{Ar}{2}]. The extended [\ion{Ne}{2}] emission in Sz~66 suggests the tentative detection of a jet.

\end{enumerate}

Binaries offer naturally controlled samples of stellar age, metallicity, and environment. Given that the primary (Sz~65) has gaps in the dust disk while the secondary  (Sz~66) does not, the cold water excess in the secondary is consistent with current theories in which gaps regulate pebble drift, and pebble drift replenishes inner-disk water through the sublimation of ice mantles. This study demonstrates the potential of wide-separation binaries for understanding the chemistry and evolution of protoplanetary disks.

\begin{acknowledgments}

This work is based on observations made with the NASA/ESA/CSA James Webb Space Telescope. The data were obtained from the Mikulski Archive for Space Telescopes at the Space Telescope Science Institute, which is operated by the Association of Universities for Research in Astronomy, Inc., under NASA contract NAS 5-03127 for JWST. These observations are associated with JWST GO Cycle 2 program ID 3034 (PI: K. Zhang). Support for J.Y. and K.Z. through this program was provided by NASA through a grant from the Space Telescope Science Institute, which is operated by the Association of Universities for Research in Astronomy, Inc., under NASA contract NAS 5-03127. I.P. acknowledges partial support from the National Aeronautics and Space Administration under agreement No.~80NSSC21K0593 for the program ``Alien Earths.'' The results reported herein benefited from collaborations and/or information exchange within NASA's Nexus for Exoplanet System Science (NExSS) research coordination network sponsored by NASA's Science Mission Directorate. N.B. and I.P. also acknowledge partial support from JWST GO Cycle 1 program ID 1621 (PI: I. Pascucci).

\end{acknowledgments}

\facility{JWST.}

\software{\texttt{Astropy} \citep{Astropy22}, \texttt{spectools-ir} \citep{Salyk22}, \texttt{emcee} \citep{Foreman13}, \texttt{Matplotlib} \citep{Matplotlib07}, \texttt{iSLAT} \citep{Jellison24, Johnson24}, \texttt{LMFIT} \citep{Newville25}.}

\appendix

\section{Continuum Subtraction}\label{App:A}

\begin{figure}[h]
\plotone{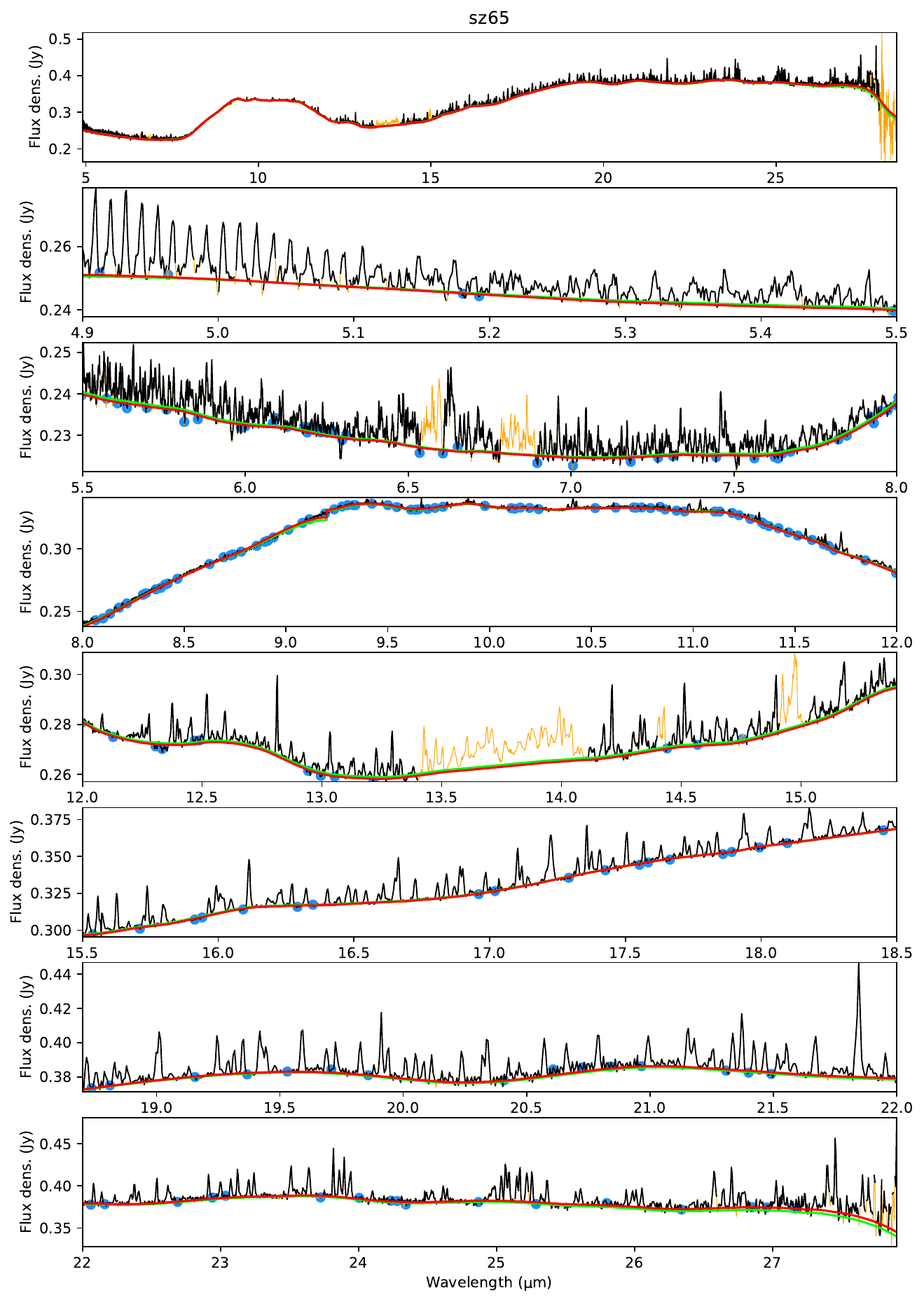}
\caption{Continuum estimation (green), continuum estimation with a calculated offset (red), and observed spectrum (black) of Sz~65. The wavelength channels excluded from the continuum subtraction and which are interpolated over are indicated in orange, and the line-free regions used for the flux offset calibration are represented by the blue dots. See Appendix B in \cite{Banzatti25} for related details.
\label{fig:sz65cont}}
\end{figure}

\begin{figure}[h]
\plotone{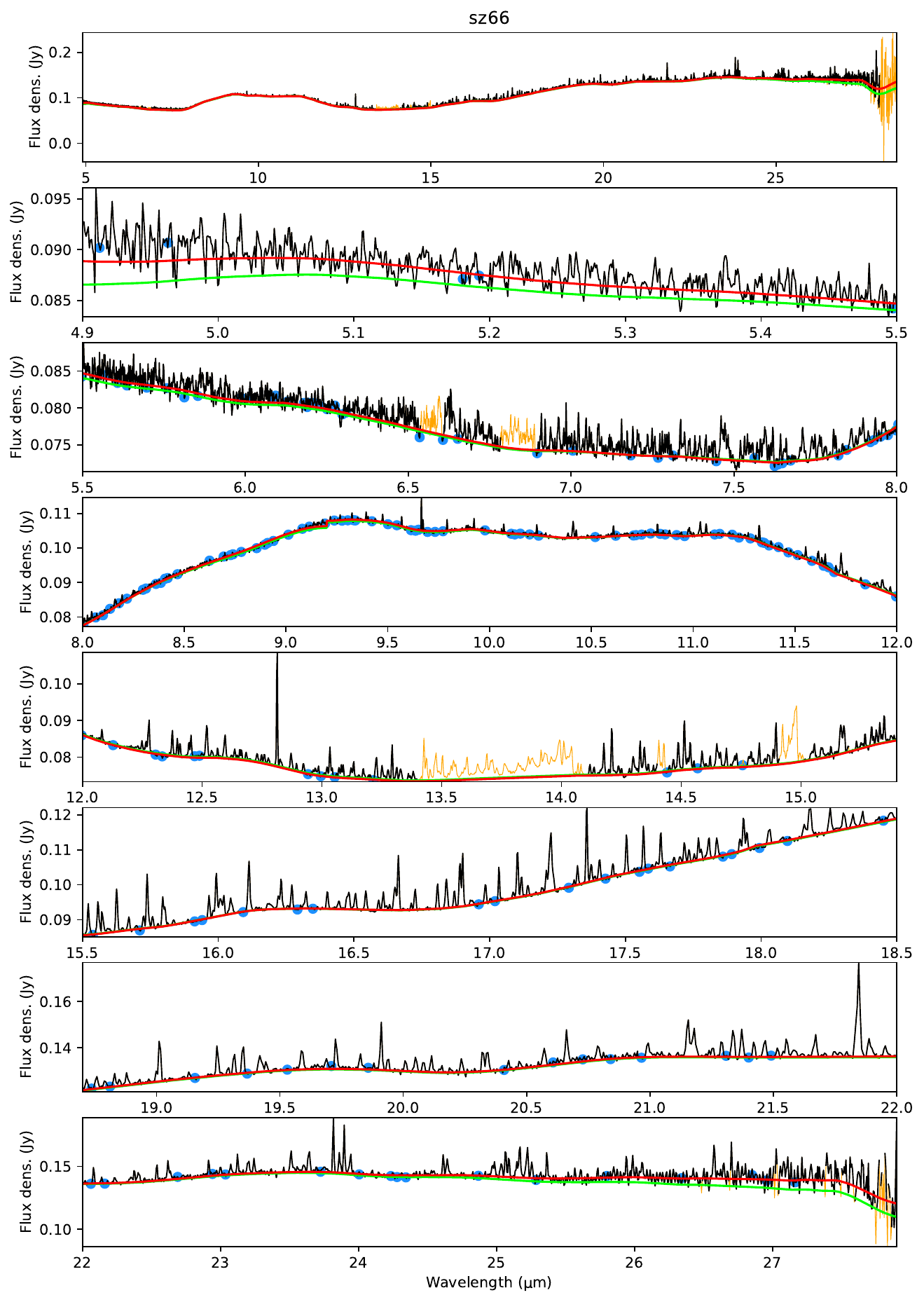}
\caption{Continuum estimation (green), continuum estimation with a calculated offset (red), and observed spectrum (black) of Sz~66. The wavelength channels excluded from the continuum subtraction and which are interpolated over are indicated in orange, and the line-free regions used for the flux offset calibration are represented by the blue dots. The estimation at 4.9--5.5\,$\mu $m might be degraded by photospheric absorption by the host star.
\label{fig:sz66cont}}
\end{figure}

We followed the continuum subtraction process presented in \cite{Pontoppidan24} and \cite{Banzatti25}, which removes dust continuum while keeping narrow gas lines. The empirical method assumes an emission-dominated spectrum and excludes regions where absorption spikes are present. We obtain the continuum by iteratively applying median filters and retaining the lower flux, followed by a second-order Savitzky–Golay filter \citep{Savitzky64}. The window sizes of the median filters are 151 and 91 for 4.9--9.2\,$\mu$m and 9.2--28.5\,$\mu$m, respectively. To avoid continuum overestimation caused by the dense clustering of lines, we exclude and interpolate over the following regions from continuum estimation: 6.538--6.603\,$\mu $m (H$_2$O), 6.784--6.89\,$\mu $m (H$_2$O), 13.4--14.1\,$\mu $m (C$_2$H$_2$ and HCN), 14.40--14.435\,$\mu $m (H$_2$O), 14.74--14.77\,$\mu $m (H$_2$O), 14.91--15.01\,$\mu $m (CO$_2$), and 28--28.5\,$\mu $m (low S/N). We selected a set of line-free regions identified in \texttt{iSLAT}, and a final offset is applied according to the fluxes at these regions. Figure \ref{fig:sz65cont} and \ref{fig:sz66cont} show the continuum estimation process of Sz~65 and Sz~66.

\section{Gaussian Noise Estimation}\label{App:B}

The noise of continuum-subtracted spectra is estimated using the Gaussian process introduced in \cite{RomeroMirza24}. The method calculates the variance $\sigma^2$ of spectral features with a correlation length smaller than the wavelength-dependent resolving power. The best-fit $\sigma$ values of Sz~65 and Sz~66 are summarized in Table \ref{tab:noise}, which are used as the estimated noise in the slab-model MCMC fitting (Section \ref{sec:models}).

\begin{deluxetable}{ccccccc}[htbp]
\tablewidth{\columnwidth}
\tabletypesize{\scriptsize}
\tablecaption{Gaussian noise of continuum-subtracted spectra of Sz~65 and Sz~66\label{tab:noise}}
\tablehead{
  \colhead{Source} &
  \colhead{12.0--13.5\,$\mu $m} &
  \colhead{13.5--15.5\,$\mu $m} &
  \colhead{15.5--18.0\,$\mu $m} &
  \colhead{18.0--21.0\,$\mu $m} &
  \colhead{21.0--24.5\,$\mu $m} &
  \colhead{24.5--27.0\,$\mu $m} \\[-1.2ex]
  \colhead{} &
  \colhead{(mJy)} & \colhead{(mJy)} & \colhead{(mJy)} &
  \colhead{(mJy)} & \colhead{(mJy)} & \colhead{(mJy)}
}
\startdata
Sz~65 & 1.45 & 1.83 & 2.47 & 3.86 & 5.49 & 6.27\\
Sz~66 & 1.26 & 1.07 & 1.53 & 1.76 & 2.88 & 4.64\\
\enddata
\end{deluxetable}

\section{Photospheric Models}\label{App:Photo}

\begin{figure*}[htbp]
\centering
\includegraphics[width=0.9\textwidth]{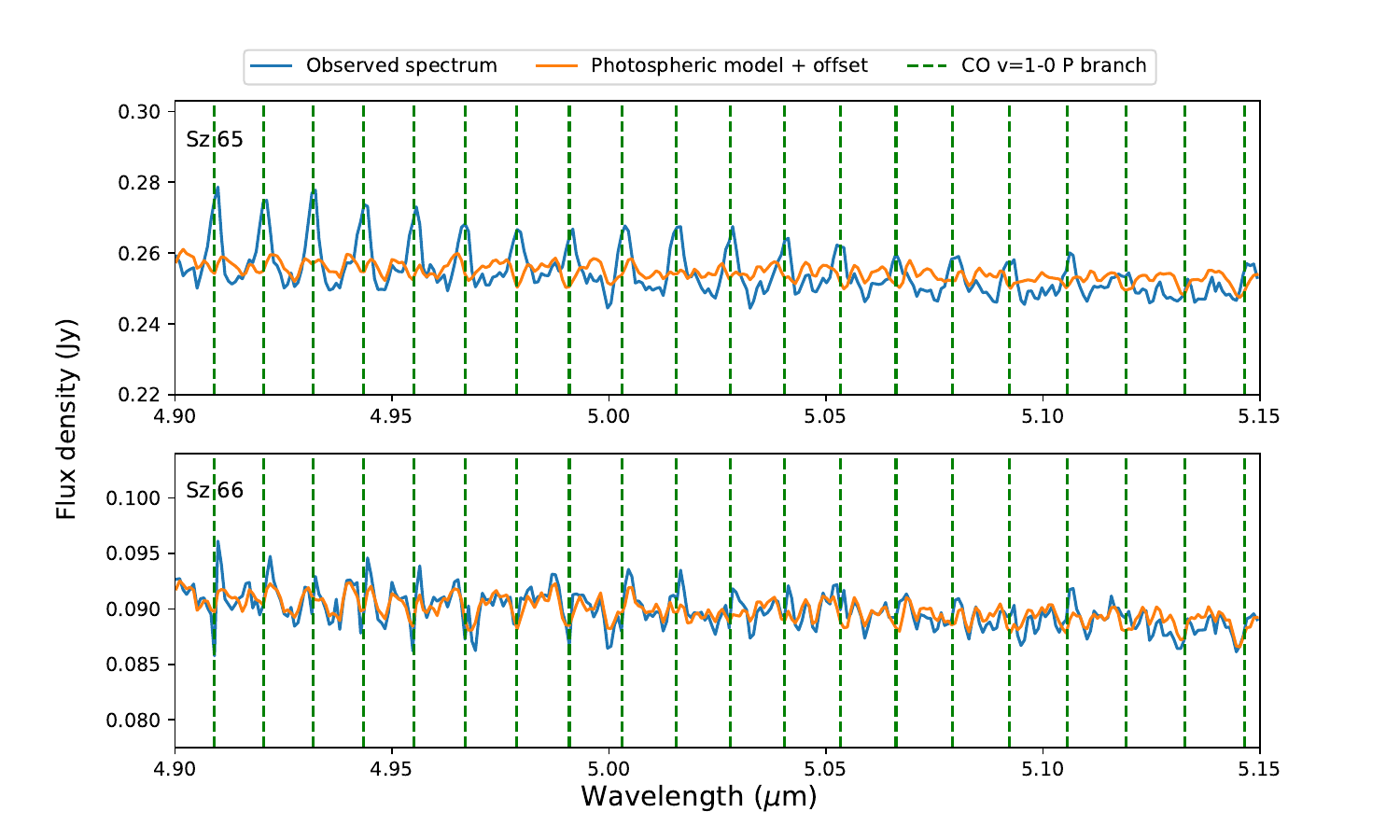}
\caption{The NewEra photospheric models \citep{Hauschildt25} of Sz~65 and Sz~66 overlaid on their observed spectra. The vertical green dashed lines indicate the CO $v=1$--0 transitions.
\label{fig:photospheric}}
\end{figure*}

We investigated the origin of the absorption features around 5\,$\mu$m exhibited by the secondary (Sz~66) by comparing the observed spectrum with photospheric models. Figure \ref{fig:photospheric} shows the NewEra photospheric models \citep{Hauschildt25} tailored to the properties ($\log g$ and $T_{\mathrm{eff}}$) of Sz~65 and Sz~66. The models shown have $T_{\mathrm{eff}}$ values of 4000\,K and 3400\,K for Sz~65 and Sz~66, respectively, and $\log g = 3.5$ for both stars. The model spectra are extinction-adjusted using $A_V$ values of 0.6 and 1.0, respectively, and convolved to a resolving power of $R = 3000$. For Sz~66, there is significant correspondence between the photospheric model and the observed spectrum, which supports a primarily photospheric origin for the absorption feature. Note that some observed lower-$J$ CO features at the short-wavelength end appear to be deeper than in the model. This may be caused by an overly broad convolution kernel, overestimated veiling, or an additional contribution from the disk atmosphere.

\section{Coldest Emission Diagnostics}\label{App:Cold}

\begin{figure}[htbp]
\centering
\includegraphics[width=0.39\textwidth]{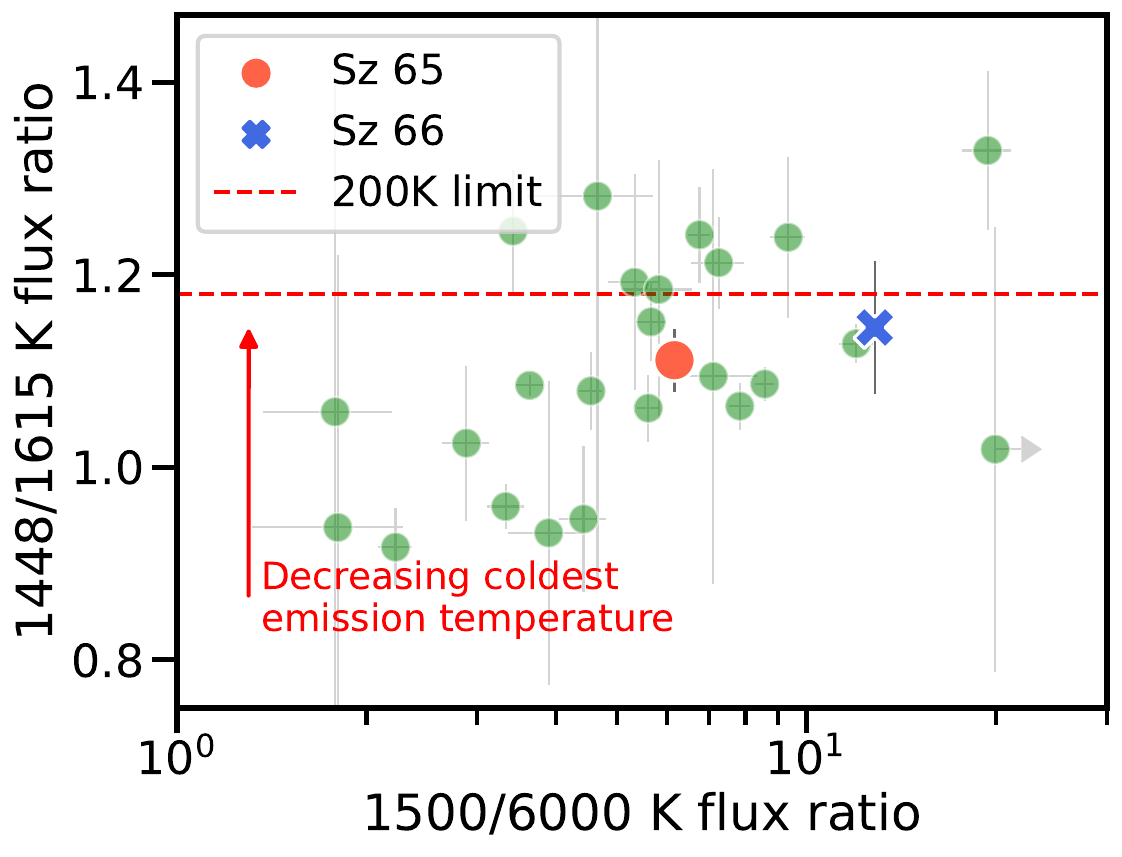}
\caption{Line diagnostic diagram showing the 1448/1615 K line ratio that traces the coldest emission and the 1500/6000 K line ratio that traces the cold water contribution. Most reference disks fall above the 200 K threshold (red dashed line), implying the coldest emitting temperatures are below 200 K. See Figure 11 in \cite{Banzatti25} for related details.
\label{fig:23um}}
\end{figure}

In addition to the water line-ratio diagnostics presented in Section \ref{sec:diag}, we also applied a two-line-ratio diagnostic using the 23.81676\,$\mu$m ($E_u=1448\,$K) and 23.89518\,$\mu$m ($E_u=1615\,$K) lines to probe the coldest H$_2$O-emitting gas (see \cite{Banzatti25} for details). As shown in Figure \ref{fig:23um}, both disks lie just below the 200 K limit, suggesting characteristic temperatures of $\sim200\,$K for the coldest emitting component, consistent with the majority of reference disks. Note that this diagnostic may have a slight dependence on column density because the two lines have different $A_\mathrm{ul}$ values (0.61 and 1.04\,s$^{-1}$, respectively). Therefore, the line ratio of the secondary (Sz~66) might be suppressed by its relatively high column density.

\section{Posterior Distributions of OH Parameters}\label{App:OH}

\begin{figure*}[htbp]
\centering
\includegraphics[width=0.9\textwidth]{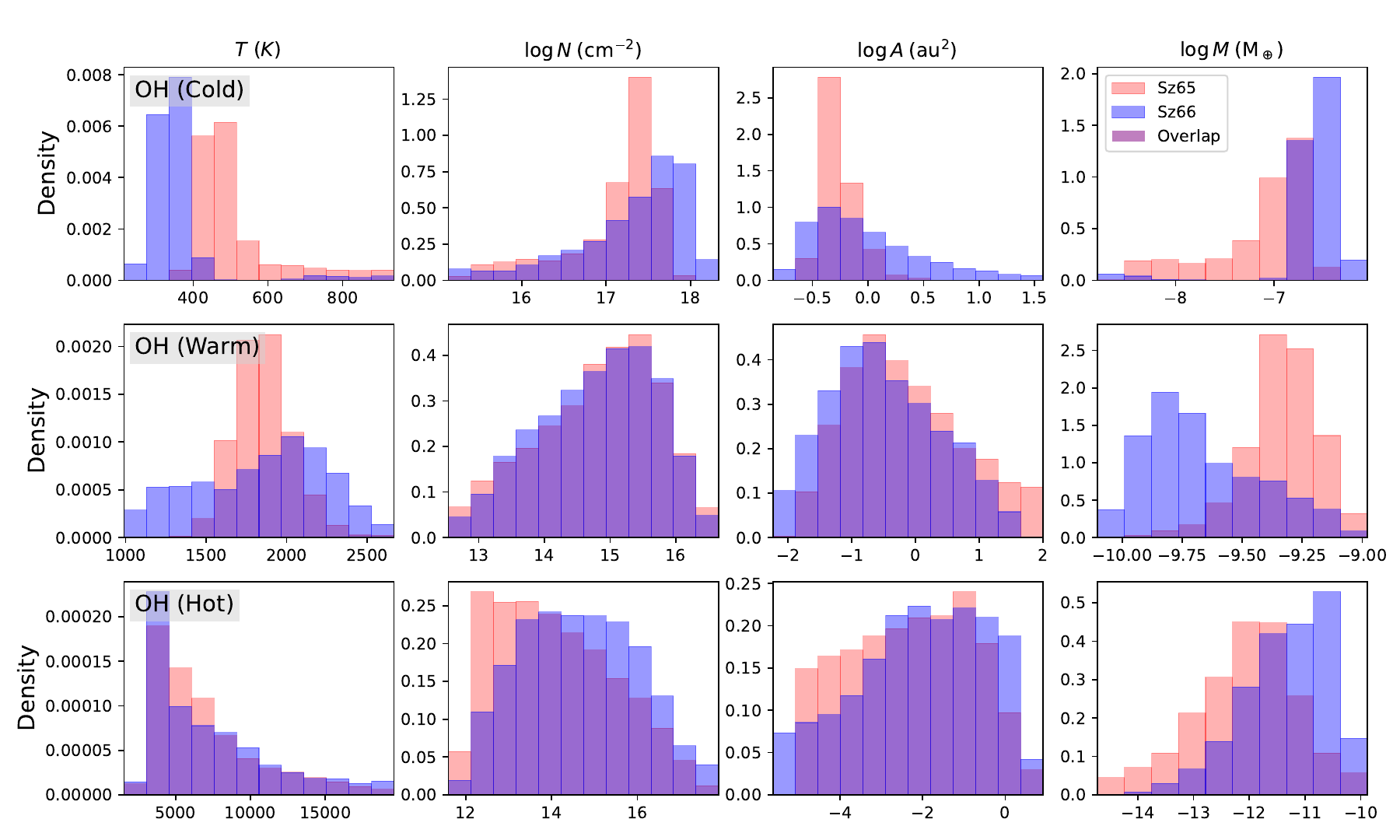}
\caption{The posterior distributions of the OH parameters for the primary Sz~65 (red) and the secondary Sz~66 (blue). Note that these parameters are unrepresentative of the thermal conditions.
\label{fig:compareHist}}
\end{figure*}

Figure \ref{fig:compareHist} presents the posterior distributions of the OH parameters fitted in Section \ref{sec:13}. The three OH components are included in the molecular model to remove OH emissions that overlap with other molecules. Because of the non-LTE effects on the OH emissions, the posterior distributions are not well constrained and not physically representative of the thermal conditions.

\section{Isotopologue Detections}\label{App:iso}



\begin{figure}[t]
\centering
\includegraphics[width=0.45\textwidth]{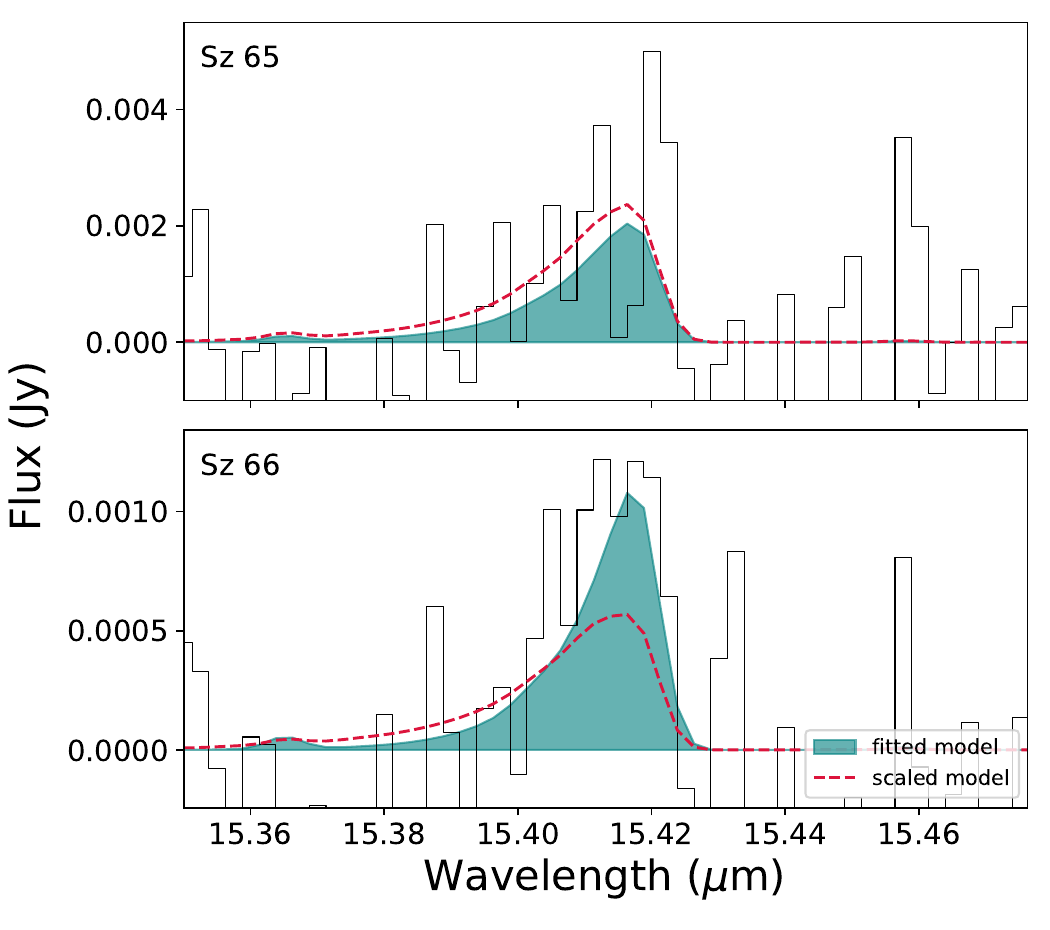}
\caption{The $^{13}$CO$_2$ models and the spectra obtained after subtracting the main emission species. The red dashed lines show the model obtained by scaling $^{12}$CO$_2$, while the cyan shades show the best-fit slab model.
\label{fig:13CO2}}
\end{figure}

\begin{figure}[t]
\centering
\includegraphics[width=0.45\textwidth]{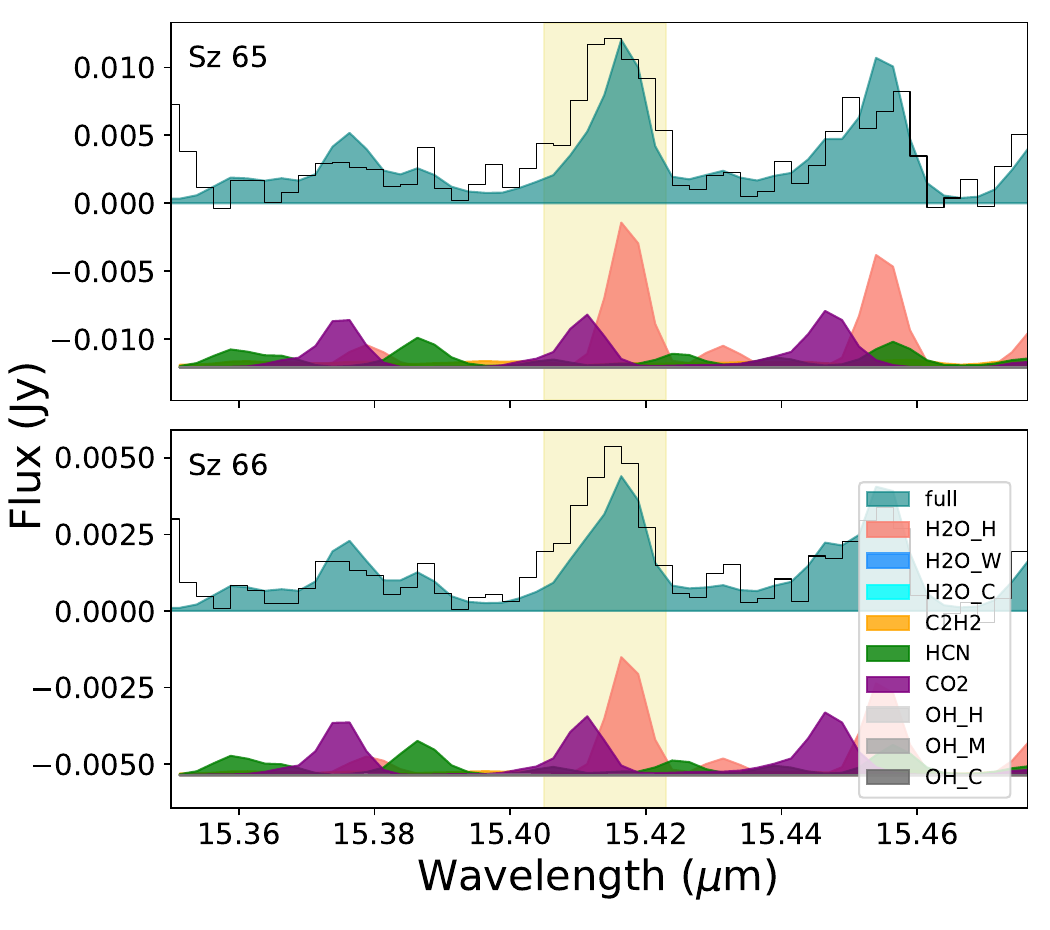}
\caption{The models of the main emission species (not including $^{13}$CO$_2$) and the continuum-subtracted spectra near the wavelength region of $^{13}$CO$_2$ emission (pale yellow).
\label{fig:13CO2full}}
\end{figure}

We inspect two isotopologues in the binary: $^{13}$CO$_2$ and $^{13}$CCH$_2$. After subtracting the model obtained from fitting 13--27\,$\mu $m, we fit the two molecules individually and evaluated the $\Delta\chi^{2}$ before and after including each molecule. The signal of $^{13}$CO$_2$ is very tentative in both disks, whereas $^{13}$CCH$_2$ is not detected.

The best-fit $^{13}$CO$_2$ parameters are listed in Table \ref{tab:model6566}. The limited residual signal and the optically thin nature of $^{13}$CO$_2$ lead to the degeneracy of multiple parameters, so the uncertainties cannot be derived with confidence, and the best-fit values are just coarse estimates. The temperature is relatively low compared to $^{12}$CO$_2$, which is commonly seen in previous detections because the traced gas is deeper in the disk \citep{Grant23, Vlasblom25}.

The best-fit mass ratios of $^{13}$CO$_2$ to $^{12}$CO$_2$ for the primary (Sz~65) and the secondary (Sz~66) are 1:50 and 1:10, respectively. The secondary's value appears elevated relative to the canonical ISM ratio of 1:68 \citep{Milam05}, but this comparison is uncertain and may be affected by optical-depth effects, as $^{13}$CO$_2$ can trace deeper layers. To further investigate the isotopologue ratio, we compare our best-fit model with a model obtained by scaling the $^{12}$CO$_2$ column density by the canonical ISM ratio to assess whether the isotopologue signals are consistent with expectations. Figure \ref{fig:13CO2} compares the best-fit $^{13}$CO$_2$ model with the scaled model, which uses 1/68 of the $^{12}$CO$_2$ column density and retains the same temperature and emitting area. The two models are very similar for the primary, whereas the best-fit model has a higher flux than the scaled model for the secondary.
However, the slab-model parameters of $^{13}$CO$_2$ are poorly constrained because its emission is relatively weak compared to the noise after model subtraction, so the best-fit results may not be representative of the isotopologue ratio. As shown in Figure \ref{fig:13CO2full}, the residual emissions of $^{13}$CO$_2$ constitute only a small fraction of the total continuum-subtracted spectra, as the H$_2$O and $^{12}$CO$_2$ emissions dominate at this wavelength.

\section{Slab-model Residuals}\label{App:underpred}

\begin{figure}[htbp]
\centering
\includegraphics[width=0.45\textwidth]{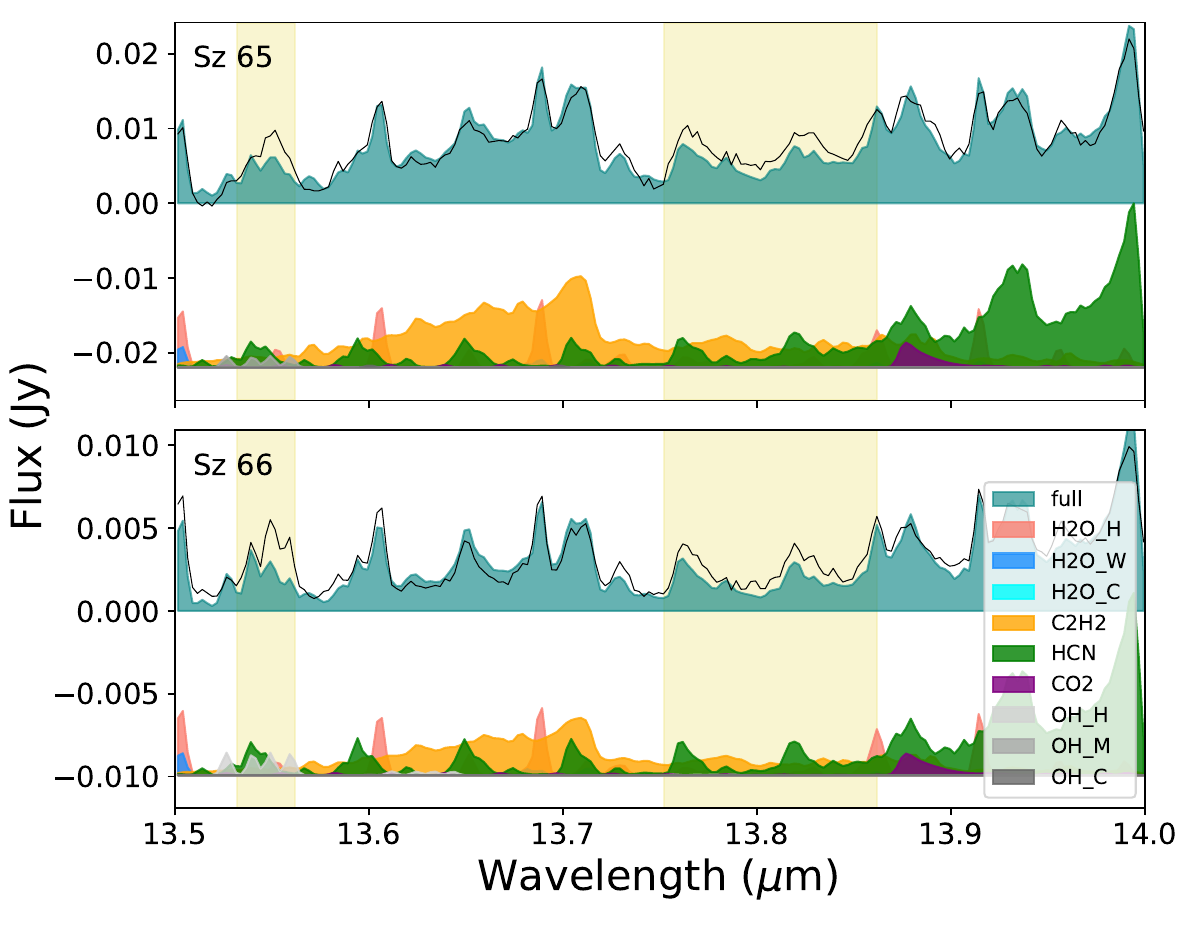}
\caption{The best-fit model and the continuum-subtracted spectra for the binary in the 13.5--14.0\,$\mu $m range. The underpredicted regions are highlighted in pale yellow.
\label{fig:under}}
\end{figure}

We identify two relatively broad residuals of our slab models at 13.55 and 13.8\,$\mu $m that appear in both disks and cannot be explained by other molecular species or atomic lines after careful examination. Figure \ref{fig:under} zooms in on this wavelength region and highlights the underpredicted features.

The underpredictions at 13.55\,$\mu $m coincide with the wavelength of an OH quadruplet. The current three-component OH model, which assumes LTE, does not reproduce this feature successfully enough. However, the OH prompt emission originating from UV dissociation of H$_2$O might explain the discrepancy. Specifically, the non-LTE prompt emission at rotational level $N=22$ contributes to the flux at 13.55\,$\mu $m. The OH prompt emission in protoplanetary disks is discussed in detail in \cite{Tabone24}. Further evidence is required to confirm the origin of this underprediction.

A broader and less prominent underprediction occurs around 13.8\,$\mu$m. This wavelength range is dominated by emission from C$_2$H$_2$ and HCN, and the residual may share a common origin with the unphysical best-fit parameters of C$_2$H$_2$ (see Table \ref{tab:model6566}). Slab-model fits excluding this region also do not yield reasonable parameters, suggesting that the underprediction is not the cause of the unphysical model parameters, but rather a consequence of the limitations of the slab model. Both the underprediction and the unphysical C$_2$H$_2$ parameters could result from imperfections in the continuum subtraction, since the continuum is interpolated across this line-dense organic region. In addition, the one-slab model assumes LTE at a uniform temperature, but organic molecules such as C$_2$H$_2$ and HCN may exhibit non-LTE excitation or temperature gradients that require multiple slab components to model (see \citealt{Tabone26}).

\bibliography{sample701}{}
\bibliographystyle{aasjournal}

\end{document}